\documentclass[twocolumn,showpacs,prb,amsmath,amssymb]{revtex4}
\usepackage{graphicx}
\usepackage{dcolumn}
\usepackage{bm}

\begin{document}
\title{Optical Phonons in Carbon Nanotubes: Kohn Anomalies, Peierls Distortions and Dynamic Effects}
\author{Stefano Piscanec$^1$}
\author{Michele Lazzeri$^2$}
\author{J. Robertson$^1$}
\author{Andrea C. Ferrari$^1$}
\email{acf26@eng.cam.ac.uk}
\author{Francesco Mauri$^2$}
\email{mauri@lmcp.jussieu.fr} \affiliation{$^1$Cambridge University,
Engineering Department, 9 JJ Thomson Avenue, Cambridge CB3 OFA, UK\\
$^2$Institut de Min\'eralogie et de Physique des Milieux
Condens\'es, Universit\'e Paris VI,  4 Place Jussieu, Paris, France
}
\date{September 23, 2006}

\begin{abstract}

We present a detailed study of the vibrational properties of
Single Wall Carbon Nanotubes (SWNTs). The phonon dispersions of
SWNTs are strongly shaped by the effects of electron-phonon
coupling. We analyze the separate contributions of curvature and
confinement. Confinement plays a major role in modifying SWNT
phonons and is often more relevant than curvature. Due to their
one-dimensional character, metallic tubes are expected to undergo
Peierls distortions (PD) at T=0K. At finite temperature, PD are no
longer present, but phonons with atomic displacements similar to
those of the PD are affected by strong Kohn anomalies (KA). We
investigate by Density Functional Theory (DFT) KA and PD in
metallic SWNTs with diameters up to 3 nm, in the electronic
temperature range from 4K to 3000 K. We then derive a set of
simple formulas accounting for all the DFT results. Finally, we
prove that the static approach, commonly used for the evaluation
of phonon frequencies in solids, fails because of the SWNTs
reduced dimensionality. The correct description of KA in metallic
SWNTs can be obtained only by using a dynamical approach, beyond
the adiabatic Born-Oppenheimer approximation, by taking into
account non-adiabatic contributions. Dynamic effects induce
significant changes in the occurrence and shape of Kohn anomalies.
We show that the SWNT Raman G peak can only be interpreted
considering the combined dynamic, curvature and confinement
effects. We assign the G$^{+}$ and G$^{-}$ peaks of metallic SWNTs
to TO (circumferential) and LO (axial) modes, respectively, the
opposite of semiconducting SWNTs.
\end{abstract}

\pacs{78.67.Ch, 63.20.Dj, 63.20.Kr, 71.15.Mb, 78.30.-j}







\maketitle

\section{Introduction}

Carbon nanotubes are at the center of nanotechnology research
\cite{ Endo1976,Iijima1991}. The determination of their structure,
phonon dispersions, and Raman spectra is a most intense area of
investigation since their discovery \cite{Rao1997, SaitoBook,
ReichBook, acfbook}. Single wall nanotubes (SWNTs) can be
described either as giant molecules or one-dimensional crystals.
Due to their reduced dimensionality, confinement effects play a
fundamental role in shaping their physical properties, such as the
metallic or semiconducting character or the electron transport
mechanisms. Optical phonons of SWNTs are extremely important.
Indeed, they contribute the most intense features in the Raman
spectra. These are commonly used to sort metallic from
semiconducting SWNTs. Electron scattering by optical phonons also
sets the ultimate limit of high field ballistic
transport~\cite{Yao2000,Park2004,Lazzeri2005,Javey2004}, due to
hot phonon generation~\cite{Lazzeri2005, LazzeriMauri2006}.
Electron-phonon coupling (EPC) is the key physical parameter
necessary to quantify the phonons interaction with electrons. In
metallic SWNTs the EPC strongly affects the phonon frequencies,
giving rise to Kohn Anomalies (KA) \cite{Piscanec2004,
Piscanec2005kirch, Piscanec2005MRS,Barnett2005, Ferrari2006} and
Peierls Distortions (PD)~\cite{Dubay2002, Dubay2003, Bohnen2004,
Connetable2005, Huang1996prb,
Huang1996ssc,Sedeki2000,Figge2001,Barnett2005}. A correct
understanding and a quantitatively precise description of
electron-phonon coupling, Kohn anomalies, and Peierls distortions
in SWNTs is then of prime scientific and technological interest.

Experimentally, the most commonly studied SWNTs have diameters
from 0.8 nm, up to 3 nm, with hundreds to thousands atoms per unit
cell. Covering such diameter range pushes the limits of
computational approaches. SWNTs optical phonons were calculated
using phonon zone folding \cite{Jishi1994, Eklund1995,
Sanchez-Portal1999, Dubay2003}, tight binding\cite{Yu1995el,
Yu1995jcp, Menon1996, Popov2006} (TB), density functional theory
(DFT), \cite{ Kurti1998, Sanchez-Portal1999, Dubay2002, Dubay2003,
Ye2004, Bohnen2004, Connetable2005} and symmetry-adapted models
\cite{Popov1999, Li2004, Milosevic2005, Mapelli1999, Popov2006}.
The simplest approach to get SWNT phonons, consists in folding
those of graphene \cite{Jishi1993cpl, Dubay2003, SaitoBook,
ReichBook}. This requires the prior calculation of graphene
phonons either by TB~\cite{Jishi1993cpl, Jishi1994,Eklund1995},
DFT~\cite{Sanchez-Portal1999, Dubay2003, Maultzsch2004prl,
Piscanec2004} or by other approaches\cite{Mapelli1999}. Phonon
zone folding (PZF) is not limited by the size of the SWNT unit
cell. However, even if SWNT electrons are well described by folded
graphene~\cite{zolyomi2004}, this is not always true for phonons.
We have shown that graphene has two KA at ${\bm \Gamma}$ and $
\mathbf{K}$~\cite{Piscanec2004}. Due to their reduced
dimensionality, metallic SWNTs have much stronger KA than
graphene\cite{Piscanec2004, Piscanec2005kirch, Piscanec2005MRS}.
Thus, folded graphene cannot give a reliable phonon dispersion of
metallic SWNTs close to the Kohn anomalies, no matter how accurate
the graphene phonons~\cite{ Maultzsch2004prl, Piscanec2004}. On
the other hand, semiconducting SWNTs cannot have KA. So, even in
this case, folded graphene may not correctly describe the phonons.

These problems can be overcome by means of TB or DFT calculations.
Indeed, the EPC effects can be taken into account in TB
models~\cite{Jiang2004, Park2004,Perebeinos2005electron}. However,
the EPC strongly depend on the TB parameterisation, and
contrasting values are found in literature~\cite{Mahan2003,
Jiang2004, Park2004,Perebeinos2005electron, Lazzeri2005}. Also,
most TB works failed to identify KA \cite{Yu1995el, Yu1995jcp,
Menon1996}. DFT directly includes EPC effects, and, being an
\emph{ab-initio} technique, does not rely on external adjustable
parameters. However, this better accuracy is obtained at the price
of a significant increase of the computational time demand, which
makes the evaluation of SWNT properties a challenge. As a result,
most DFT works focus on extremely small tubes
\cite{Connetable2005, Bohnen2004} and/or achiral tubes \cite{
Sanchez-Portal1999, Dubay2002, Dubay2003}. Finally, in all the
approaches used to compute phonon frequencies, the time dependent
nature of phonons is neglected. This approximation, which results
in a \emph{static} description of the atomic vibration, is usually
legitimate in tri-dimensional crystals \cite{Baroni2001,
Degironcoli1995}, but, as we will show, does not necessarily hold
for one-dimensional systems.

In this paper we show that SWNTs are a major exception, and that a
correct description of their optical phonons can only be achieved
by carefully considering the dynamical nature of phonons. We first
analyze the static case and single out the effects of curvature
and confinement. Confinement plays a major role in shaping SWNT
phonons and is often more relevant than curvature. We present an
electronic zone folding method allowing the static DFT calculation
of confinement effects on phonon-dispersions and EPC of SWNTs of
any diameter and chirality, and for any electronic temperature
($T_{\rm e}$). We investigate KA and PD in metallic SWNTs with
diameters up to 3 nm and for $T_{\rm e}$=4 to 3000 K. We present a
simple analytic model exactly accounting for all the static DFT
results. Finally, we prove that dynamic effects induce significant
changes in KA occurrence and shape. We show that the SWNT Raman G
peak can only be interpreted considering the combined dynamic,
curvature and confinement effects. We assign the G$^{+}$ and
G$^{-}$ peaks of metallic SWNTs to TO (circumferential) and LO
(axial) modes, the opposite of semiconducting SWNTs. A family
dependence of the G peak position in semiconducting tubes is
observed.

This paper is organized as follows. In Section II we introduce the
background concepts and definitions. Section III presents our EPC
calculations. Section IV and V present respectively the study of
the effect of pure confinement within the static and the dynamic
approach. Section VI highlights the influence of curvature on the
SWNT phonons. Finally, Section VII assigns the G$^+$ and G$^-$
Raman peaks of SWNTs.

\section{Background}

Before presenting our numerical and analytic results, it is
necessary to introduce and discuss a set of background concepts and
definitions, which will be used thorough the paper, in order to
highlight the differences between our approach and others in
literature.

\subsection{SWNTs unit cell and graphene zone folding}

SWNTs are usually identified by means of their chiral indices $n$
and $m$\cite{SaitoBook, ReichBook}. The determination of the
chiral indices is done by unrolling the unit cell of the tube and
comparing it to an infinite graphene sheet. The unrolled unit cell
is a rectangular sheet of graphene that can be described by means
of the vectors $\bf C_{\rm h}$ and $\bf T$. $\bf C_{\rm h}$ is
known as the chiral vector and has the same length as the
circumference of the tube. $\bf T$ is known as the translation
vector and defines the translational symmetry of the tube. The
crystal lattice of graphene can be defined by two vectors,
$\textbf{a}_1$ and $\textbf{a}_2$, forming a $\pi/3$ angle. $\bf
C_{\rm h}$ and $\bf T$ can be projected on $\textbf{a}_1$ and
$\textbf{a}_2$~\cite{SaitoBook,ReichBook}:
\begin{equation}\label{ChT}
{\bf C}_{\rm h}=n {\bf a}_1+m {\bf a}_2 ~~~ {\bf
T}=\frac{(2m+n){\bf a}_1}{d_{\rm R}}-\frac{(2n+m){\bf a}_2}{d_{\rm
R}},
\end{equation}
where $n$, $m$ are defined as the chiral indices and $d_{\rm
R}=\gcd (2n+m, 2m+n)$ (gcd= greatest common divisor).

Wavevectors in nanotubes have to be commensurate to the tube
circumference. They can be represented on the unrolled unit cell
of the tube, and are in the form:
\begin{equation}
\textbf{k} = \textbf{k}_\bot+\textbf{k}_\|
=\mu\textbf{K}_1+\nu\textbf{K}_2, \label{K=Kpar+Kper}
\end{equation}
where $\textbf{K}_2$ is parallel to $\bf T$, its modulus $
K_2=2\pi/T$ correspond to the length of the tube's Brillouin zone
(BZ), and $\nu$ can assume any real value; $\textbf{K}_1$ is
perpendicular to $\bf T$, and $\mu\textbf{K}_1$ must be
commensurate to the SWNT circumference. Thus $K_1=2\pi/C_{\rm h}$,
$\mu\in\mathbb{N}$, and $\mu \in [N_c/2-1,N_c-1]$, $N_c$ being the
number of unit cells of graphene contained in the unrolled unit
cell of the tube. The number of nodes of a wavefunction along the
tube circumference is given by $i=2\mu$~\cite{Saito1992,
Mintmire1992, SaitoBook, ReichBook}.

Fig.~\ref{3FoldingTubo(4,4)} plots the wavevectors commensurate to
a $(4,4)$ tube on the reciprocal space of graphene, showing that
they are located on lines oriented along $\textbf{K}_2$, spaced by
$\textbf{K}_1$ and including the $\bm \Gamma$ point~\cite{
Saito1992, Mintmire1992, SaitoBook, ReichBook}. For the generic
$(n,m)$ tube, we refer to this ensemble of cutting lines as
$ZF_{(n,m)}$. The procedure of using wavevectors of graphene to
describe wavevectors of a SWNT is known as Zone Folding (ZF).
Using the correspondence set by ZF between graphene and tube
wavevectors, it is possible to use the electrons and phonons of
graphene to obtain the band structure or the phonon dispersions of
SWNTs~\cite{Saito1992, Mintmire1992, SaitoBook, ReichBook}.

For clarity, in the rest of this paper we will distinguish the
electron and phonon wavevectors by labeling them respectively with
the symbols \textbf{k} and \textbf{q}.

\begin{figure}[!tbh]
\centerline{\includegraphics[width=80mm]{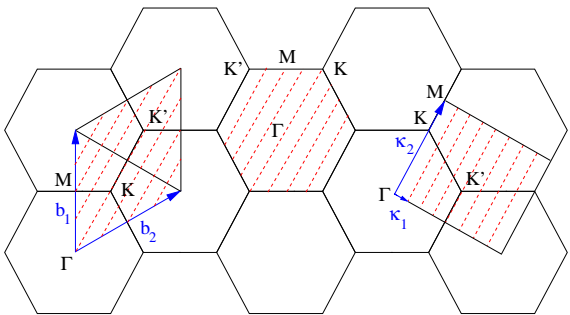}}
\caption{(color on-line) Three equivalent representations of the
zone-folding lines for a $(4,4)$ tube, $ZF_{(4,4)}$, in the
reciprocal space of graphene. Left: plot of $ZF_{(4,4)}$ in the
hcp BZ of graphene; the primitive vectors ${\bf b}_1$ and ${\bf
b}_2$ are plotted in blue, the lines of $ZF_{(4,4)}$ are plotted
in red. Center: plot of $ZF_{(4,4)}$ in the hexagonal
representation of the graphene BZ. On the right: plot of
$ZF_{(4,4)}$ emphasizing the vectors {\rm \bf K}$_1$ and {\rm \bf
K}$_2$. The length of {\rm \bf K}$_2$ corresponds to the length of
the $(4,4)$ tube BZ.} \label{3FoldingTubo(4,4)}
\end{figure}

\subsection{LO-TO phonons in graphene, graphite and nanotubes}

The analysis of the SWNT Raman G band shows that the phonons
deriving from the $\bm \Gamma-E_{2g}$ mode of graphene, of $A$
symmetry in SWNTs, provide the dominant contribution to the G$^+$
and G$^-$ Raman peaks~\cite{Jorio2003, Maultzsch2002prb}. Thus, in
this paper we focus on these modes and their phonon branches.

In literature, these phonons are often labeled using different
criteria~\cite{SaitoBook, ReichBook, Dubay2003, Jorio2000,
Reich2001, Saito2001}: (i) their symmetry\cite{SaitoBook, ReichBook,
Jorio2000}, (ii) the direction of the atomic displacements with
respect to the SWNT axis, radius or circumference\cite{Dubay2003},
(iii) the longitudinal or transversal character of the graphene
phonon branch from which they originate\cite{Reich2001}, (iiii) the
longitudinal or transversal character relative to the
SWNT\cite{Saito2001}.

In this paper the following definition is used: a mode is
\textit{longitudinal} if $\bm \epsilon \| {\rm \bf q}$ and
\emph{transverse} if $\bm \epsilon \bot {\rm \bf q}$; where $\bm
\epsilon$ is the phonon polarisation (i.e. the direction of the
atomic displacements) and ${\rm \bf q}$ its wavevector, which also
defines the direction of the phonon propagation.

In SWNTs, phonons can propagate only along the
axis. As a result, modes with atomic displacements parallel to the
axis are longitudinal, and those with atomic displacements
perpendicular to the axis (i.e. along the circumference) are
transversal.

In achiral SWNTs, symmetry forces the displacements of modes
originating from graphene $\bm \Gamma-E_{2g}$ to be either
parallel or perpendicular to the axis~\cite{SaitoBook, ReichBook}.
Thus, in armchair and zigzag SWNTs these are \textit{exactly} LO
and TO.

In chiral SWNTs, phonon polarizations are expected to be
influenced by chiral angle and C-C bonds
orientation~\cite{Reich2001}. Thus, the modes originating from the
graphene $\bm \Gamma-E_{2g}$ should have only a prevalently-LO or
a prevalently-TO character. Indeed, Ref.~\onlinecite{Reich2001}
proposed the LO/TO classification to be meaningless in chiral
SWNTs. However, in Section VI we will show that for any SWNTs with
diameter $> 0.8$~nm, these modes are almost exactly LO and TO. We
thus consider the LO/TO classification meaningful and convenient,
also because it avoids different labeling in chiral and achiral
SWNTs.

\subsection{Kohn anomalies and Peierls distortions}

\begin{figure}[!tp]
\centerline{\includegraphics[width=60mm]{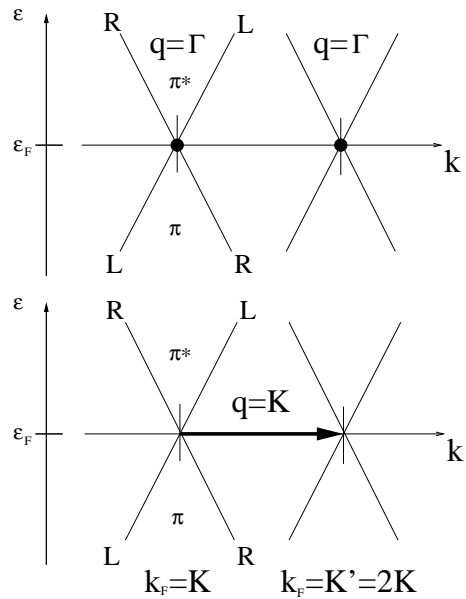}}
\caption{Geometrical condition for the KA onset in metallic SWNTs:
for q=0 a point of the Fermi surface is connected to itself. This
corresponds to a KA at $\bm \Gamma$. $q=K$ connects the two
distinct Fermi points {\rm \bf K} and \textbf{K'}, giving a second
KA. $\epsilon_{\rm F}$ is the Fermi energy. We label the band with
positive slope $L$, and that with negative slope $R$. In electron
transport, this notation indicates bands where electrons enter the
tube respectively from the left and right electrode.} \label{2kf}
\end{figure}

In a metal, for certain phonons with a wavevector connecting two
points of the Fermi surface it is possible to have an abrupt
change of the electronic screening of the atomic vibrations. This
results in a sudden softening of the phonon-frequencies, which is
called Kohn anomaly\cite{Kohn1959,AshcroftBook}. In metallic SWNTs
the Fermi surface consists of two points only. Thus, KA can occur
only for phonons with $q=0$ or $q$ connecting the two Fermi points
(Fig.~\ref{2kf}). In graphene, a necessary condition for KA is a
significant non-zero EPC (between electrons near the Fermi energy)
for phonons at ${\bf q=}{\bm \Gamma}$ or ${\bf q =
K}$~\cite{Piscanec2004}. Due to their reduced dimensionality,
metallic SWNTs are expected to have stronger KA than graphite and
graphene~\cite{Piscanec2004, Barnett2005, Bohnen2004,
Connetable2005,Piscanec2005kirch, Piscanec2005MRS}, for the
corresponding phonons.

One dimensional metallic systems are predicted to be unstable at
T=0~K\cite{Peierls}. In fact, in 1D metals it is always possible
to find a lattice distortion that (i) opens a gap in the
electronic structure, and (ii) such that the energy gained by gap
opening compensates the strain induced by the deformation. This is
called Peierls distortion\cite{Peierls}. The possible presence of
PD in SWNTs was argued as soon as their metallic behavior was
predicted\cite{Mintmire1992, Saito1992}. Their geometry was
investigated by several authors with contrasting results. It was
proposed that they follow the atomic displacement patterns of LA
phonons\cite{Huang1996prb, Huang1996ssc}, of Radial Breathing
Modes (RBM)\cite{Sedeki2000}, of solitwistons\cite{Figge2001}, or
of optical phonons with $q =  \Gamma$ or $q = 2k_{\rm
F}$~\cite{Dubay2002, Dubay2003, Bohnen2004, Connetable2005,
Barnett2005, Piscanec2005kirch}, where $k_{\rm F}$ is the
wave-vector for which the electron gap is null. Note that the
condition $q = 2k_{\rm F}$ is equivalent to $q =K$ in
Fig.~\ref{2kf}.

For $T>0$K, due to thermal excitation, states above the Fermi
energy are populated. Thus, increasing the temperature reduces the
energy gained by gap opening. PD are then stable only below a
critical temperature $T_{\rm PD}$. This temperature is also known
as the metal-semiconductor transition temperature, and its
determination is crucial to understand the interplay between the
possible onset of superconductivity and
PD~\cite{Benedict1995,Tang2001,Bohnen2004,Connetable2005,
Barnett2005}.

For $T>T_{\rm PD}$, the energy gain achieved by gap opening is not
sufficient to compensate the elastic strain, so the lattice will
not undergo a permanent distortion. However, phonon modes having a
pattern of atomic displacements corresponding to the PD can show
an important softening. This effect is proportional to the energy
gained by gap opening, and inversely proportional to the
temperature. This mechanism, described by some authors as a
Peierls-like mechanism\cite{Dubay2002, Dubay2003}, is, in fact, a
particular case of Kohn Anomaly\cite{Kohn1959}.

\subsection{Temperature Effects}

Let us call $T_{\rm e}$ and $T_{\rm i}$ the electronic and the
ionic temperature. $T_{\rm i}$ corresponds to the energy
associated to the atoms vibrations around their equilibrium
positions. $T_{\rm e}$ fixes the electronic states population
close to the Fermi energy $\epsilon_{\rm F}$:
\begin{equation}
f_{\textbf{k},n}=\frac{1}{e^{(\epsilon_{\textbf{k},n}-\epsilon_{\rm
F})/k_{\rm B}T_{\rm e}}+1}, \label{FermiDirac}
\end{equation}
where $f_{\textbf{k},n}$ is the Fermi-Dirac occupation of the
$n^{th}$ electronic band at \textbf{k}, and $k_{\rm B}$ is the
Boltzmann's constant.

In metals, a finite electronic temperature always changes the
electronic states occupation, while in materials with an
electronic gap $E_{\rm g}$, changes of the electronic occupancies
are relevant only for $k_{\rm B}T_{\rm e}>E_{\rm g}$.

The Peierls-distortion temperature $T_{\rm PD}$ depends only on
the electronic structure of the system, and not on the lattice
thermal energy. Thus, Peierls distortions are features of the
phonon dispersions which directly depend on $T_{\rm e}$. On the
other hand, phonons can have a temperature dependence due to
anharmonicity~\cite{AshcroftBook}, which is determined by $T_{\rm
i}$.

It is thus essential to distinguish the different effects of
$T_{\rm e}$ and $T_{\rm i}$ on the SWNT phonons. At thermal
equilibrium, $T_{\rm i}=T_{\rm e}=T_{\rm S}$, where $T_{\rm S}$ is
the sample temperature. Thus, the correct phonon frequencies can
be recovered only by taking simultaneously into account the
contributions of Kohn anomalies and anharmonicity. The calculation
of the anharmonic effects will be discussed
elsewhere\cite{mauriump2006}. To a first approximation, the
anharmonic contribution is similar for phonons of similar energy.
On the contrary, a key result of this paper will be to show that
$T_{\rm e}$ \emph{selectively} acts only on the TO and LO phonon
branches in metallic SWNTs, see Section IV and V.

DFT, in its original formulation, is a ground state theory, thus
$T_{\rm e}=0 K$. DFT can be extended to include a finite
\emph{real} $T_{\rm e}$\cite{Mermin1965}. This is done using the
Mermin potential~\cite{Mermin1965} and populating the electronic
states with the Fermi-Dirac distribution, Eq.~\ref{FermiDirac}.
Because of the presence of KA and PD, the phonon frequencies of
metallic SWNTs depend on $T_{\rm e}$ and DFT calculation should
always be done using the Mermin potential.

\subsection{The dynamical matrix}

We call $\Theta_{\rm \bf q}$~\cite{ NotaNotazione} the dynamical
matrix projected on the phonons normal coordinates. Within time
dependent perturbation theory, for a SWNT this is defined as (see,
e.g., Eqs. 4.17a, 4.23 in Ref.~\onlinecite{Allen1980}):
\begin{eqnarray}
\Theta_{q} &=& \frac{2T}{2\pi} \sum_{m,n}\int_{BZ} \frac{|D_{(
 k+q)n, km}|^2[f_{k,m}-f_{k + q,n}]}
{\epsilon_{k,m}-\epsilon_{
k+q,n}+\hbar\omega_{q}+i\gamma} dk \nonumber \\
&&-\int\Delta n^*_{q}({\rm \bf r}) K({\rm \bf r},{\rm \bf r'})
\Delta n_{q}({\rm \bf r'})~d{\rm \bf r} d{\rm \bf r}' \nonumber
\\&&+\int n({\rm \bf r}) \Delta^2V^{\rm b}({\rm \bf r})~d{\rm \bf r},
\label{DynMat on ph normal coord dynamic}
\end{eqnarray}
where $2\pi/T$ is the length of the tube BZ, $\sum_{m,n}$ is a sum
on all the possible electronic transitions, $\int_{BZ}dk$ is an
integral over the one dimensional tube BZ; $\epsilon_{k,n}$ is the
energy of the electronic Bloch eigenstate with wavevector $k$ and
band index $n$; $\gamma$ is a small real constant; $n({\rm \bf
r})$ is the charge density; $K({\rm \bf r,r}')=\delta^2 E_{\rm
Hxc}[n]/\delta n({\rm \bf r})\delta n({\rm \bf r}')$, where
$E_{\rm Hxc}[n]$ is the Hartree and exchange-correlation
functional, and $\Delta^2V^{\rm b}$ is the second derivative of
the bare (purely ionic) potential; $D_{(k+q)n,km}$ is the
electron-phonon coupling matrix element:
\begin{equation}
D_{(k+q)n,km} = \langle k+q,n| \Delta V_{q}{[\Delta n_{q}]}
|k,m\rangle, \label{definition of G}
\end{equation}
where $\Delta V_{q}$ and $\Delta n_{q}$ are the derivatives of
Kohn-Sham potential and charge density with respect to
displacement along the phonon normal coordinate; $|k,n\rangle$ is
the electronic Bloch eigenstate of wavevector $k$ and band $n$.
The phonon frequencies $\omega_q$ are derived from $\Theta_{q}$
as~\cite{ Allen1980}:
\begin{eqnarray}
\omega_{q}&=&\Re e\bigg\{\sqrt{\frac{\Theta_{q}}{M}}\bigg\},
 \label{omega=sqrt(Dq/M)}
\end{eqnarray}
where M is the atomic mass of carbon.

It is important to note that Eq.~\ref{DynMat on ph normal coord
dynamic} introduces a dependence of the dynamical matrix on the
term $\hbar\omega_{q}+i\gamma$. This is a direct consequence of
the perturbative approach used to derive Eq.~\ref{DynMat on ph
normal coord dynamic}, where the phonon is described as a
time-dependent perturbation of the system~\cite{ Allen1980}. This
accounts for the dynamic nature of vibrations. In the rest of the
paper, we will refer to the inclusion of the term
$\hbar\omega_{q}+i\gamma$ in the expression of the dynamical
matrix as "the inclusion of", or "the description of"
\emph{dynamic effects} (DE). This definition is introduced in
opposition to the \emph{static approximation}, which is described
in Section II-E.2.

\subsubsection{Dynamic effects}

Dynamic effects induce a correction in the denominators of terms
proportional to the EPC in Eq.~\ref{DynMat on ph normal coord
dynamic}. In materials where the electronic gap is larger than the
phonon energy, the dynamic effects are negligible. In metals,
dynamic effects are negligible if the phonon energy is smaller
than the electronic smearing for which the phonon calculations
converge. In general, this is true for tri-dimensional
metals~\cite{Degironcoli1995, Baroni2001}. Thus, in this case the
contribution of the dynamic effects to the dynamical matrix is
negligible. That is why dynamic effects are not included in the
most common \emph{ab-initio} codes for phonon
calculations~\cite{PWmanual, CPMD}. However, we will show that
this is not necessarily true for systems with reduced
dimensionality. For example, as explained in the previous Section,
one dimensional metallic systems undergo Peierls distortions, the
phonon frequencies depend on the electronic temperature, and the
phonon energy can be of the same order, or even larger than the
electronic temperature. In this case, it is conceptually wrong to
neglect the dynamic effects \emph{a priori}.

Since metallic SWNTs can undergo Peierls distortions, the role of
the dynamic effects on phonon frequency has to be carefully
considered and investigated. This will be done in Section V.

\subsubsection{The Static Approximation}

The Born-Oppenheimer~\cite{Born1927}, or adiabatic, or static
approximation is equivalent to dropping the phonon energy
$\hbar\omega_q$ in Eq.~\ref{DynMat on ph normal coord
dynamic}~\cite{Sham1974}. Implementation of the static
approximation results in what we call static DFPT. To the best of
our knowledge, all existing DFPT codes implement static equations.
Static DFPT gives the same results as other static approaches,
such as frozen phonons~\cite{Yin1982}, where phonons are obtained
evaluating the energy associated to a static lattice distortion.
The expression of $\Theta_{q}$ within the static approximation is
obtained from Eq.~\ref{DynMat on ph normal coord dynamic} by
setting to zero the phonon energy $\omega_{q}$ and the term
$i\gamma$:
\begin{eqnarray}
\Theta_{q} &=& \frac{2T}{2\pi} \sum_{m,n} \int_{FBZ} \frac{f_{
k,m}-f_{k+q, n}}
{\epsilon_{k,m}-\epsilon_{k+q,n}}|D_{(k+q)n,km}|^2~dk
  \nonumber \\
&&-\int\Delta n^*_{q}({\rm \bf r}) K({\rm \bf r},{\rm \bf r'})
\Delta n_{q}({\rm \bf r'})~d\textbf{r} d\textbf{r}' \nonumber \\
&&+\int n({\rm \bf r}) \Delta^2V^{\rm b}({\rm \bf r})~d\textbf{r}.
\label{DynMat on ph normal coord}
\end{eqnarray}

Unlike in the dynamic case, within the static approximation
$\Theta_{q}$ is real. Thus, the phonon frequencies $\omega_{q}$
are simply given by:
\begin{equation}
\omega_{q}=\sqrt{\frac{\Theta_{q}}{M}}
 \label{omega=sqrt(Dq/M)_static}
\end{equation}

\subsubsection{Non analytic terms}

For states close to the Fermi energy, the denominator in
Eq.~\ref{DynMat on ph normal coord} goes to zero, resulting in the
possible presence of singularities in the phonon dispersion, as we
have shown in graphene and graphite~\cite{Piscanec2004}. Similar
singularities occur in the real part of Eq.~\ref{DynMat on ph
normal coord dynamic} when the energy of the phonon matches an
electronic transition (see, e.g., Eq.~D.17 in
Ref.~\onlinecite{MartinBook2004}). To investigate such
singularities in SWNTs it is convenient to split the dynamical
matrix into an analytic and a non-analytic component:
\begin{equation}
\Theta_{q}=\Theta_{q}^{\rm an}+{\tilde{\Theta}_{q}}.
\label{D=D_an+D_non-an}
\end{equation}

The non analytic terms in the dynamical matrix of metallic tubes can
be obtained by restricting the integral in the first term of
Eqs.~\ref{DynMat on ph normal coord dynamic} and~\ref{DynMat on ph
normal coord} to a set of k-points very close to the Fermi vector,
and by limiting the sum over the electronic bands, taking into
account only transitions between the bands crossing the Fermi point.
Labeling with $L$ the electronic band with a positive slope in
Fig.~\ref{2kf}, and $R$ that with negative slope, from
Eqs.~\ref{DynMat on ph normal coord dynamic} and~\ref{DynMat on ph
normal coord} we get:
\begin{eqnarray}
\tilde{\Theta}_{q}&=&\frac{2A_{\Gamma/K}T}{2\pi}
\sum_{m,n=L,R}\int_{-\bar k}^{\bar k}|D_{(K+k'+q)n,(K+
k')m}|^2 \nonumber \\
&&\frac{f_{K+k',m}-f_{K+k'+q,n}}{\epsilon_{K+k',m}-\epsilon_{K+k'+q,n}+\hbar\omega_q+i\gamma}dk',
 \label{Dtilde dynamic}
\end{eqnarray}
if dynamic effects are included. $\bar k$ has a small but finite
value. In this case, divergencies occur when $\epsilon_{
K+k',m}-\epsilon_{K+k'+q,n}\pm\hbar\omega_q=0$.

Within the static approximation we get:
\begin{eqnarray}
\tilde{\Theta}_{q}&=&\frac{2A_{\Gamma/
K}T}{2\pi}\sum_{m,n=L,R}\int_{-\bar k}^{\bar k} \frac{f_{
K+k',m}-f_{K+k'+ q,n}} {\epsilon_{
K+k',m}-\epsilon_{K+k'+q,n}} \nonumber \\
&& |D_{(K+k+q)n,(K+k')m}|^2
 dk'.
 \label{Dtilde}
\end{eqnarray}
Here divergencies occur when
$\epsilon_{K+k',m}=\epsilon_{K+k'+q,n}$.

For both Eq.~\ref{Dtilde dynamic} and Eq.~\ref{Dtilde},
divergences occur for $q\sim 0$ or $q\sim K$ (Fig.~\ref{2kf}). In
the first case, $A_{\Gamma}=2$, in the second case $A_{K}=1$. Note
that the distinction between the analytical and non-analytical
part of $\Theta_q$ is \emph{operative}, not \emph{physical}. In
fact, $\tilde{\Theta}_q$ contains \emph{all} the divergent terms
of $\Theta_q$ but \emph{not only} them. The cut between
$\tilde{\Theta}_q$ and $\Theta^{\rm an}_q$ depends on $\bar{k}$.

\subsubsection{Curvature and Confinement}

The differences between graphene and a SWNT can be described in
terms of \textit{curvature} and \textit{confinement}.
\textit{Curvature effects} are due to the distortion of the C-C
bonds in the SWNT geometry and in a change of the bond character
with respect to planar graphene. \textit{Confinement effects} stem
from the reduced dimensionality and are due to the quantization of
the electronic wavefunctions along the SWNT circumference.

In this paper we concentrate on the description of the effects of
confinement and curvature on the dynamical matrix and on the
phonon frequencies of SWNTs. Considering the effects of curvature
and confinement as perturbations on the dynamical matrix of
graphene $\hat{\Theta}_{\rm G}$, at first-order:
\begin{equation}
\hat{\Theta}_{\rm T}=\hat{\Theta}_{\rm G}+\hat{\Theta}_{\rm
curv}+\hat{\Theta}_{\rm conf},
 \label{Dtubo1}
\end{equation}
where $\hat{\Theta}_{\rm T}$ is the dynamical matrix of a SWNT and
$\hat{\Theta}_{\rm curv}$ and $\hat{\Theta}_{\rm conf}$ account
for curvature and confinement. The effects of pure confinement
will be calculated in Section IV and V. The effects of curvature
will be evaluated in Section VI.

\section{Electron-phonon coupling}

In Ref.~\onlinecite{Lazzeri2005} we have shown that the EPC of SWNT
of arbitrary chirality is weakly affected by the tube curvature, and
that it can be computed by folding the graphene EPC. Neglecting
curvature, the EPC of a $(n,m)$ SWNT is related to that of graphene
by the ratio of the unit cells:
\begin{equation}
|D^2||{\bf C}_{\rm h} \times {\bf T}|= |\tilde D^2|
\frac{a_0^2\sqrt{3}}{2} \label{G^2tubo and G^graph}
\end{equation}
where $D$ is the EPC of a $(n,m)$ SWNT \emph; $\tilde D$ is the
graphene EPC; ${\bf C}_{\rm h} \times {\bf T}$ the area of the
unrolled tube unit cell; $a_0$ is the lattice parameter of
graphene, and $a_0^2\sqrt{3}/2$ is the graphene unit cell area.
This formula is a simple consequence of electron and phonon
normalization in the two different unit cells. Its derivation is
presented in Appendix A.

Since ${\bf T}$ and ${\bf C}_{\rm h}$ are perpendicular,
$|\textbf{C}_{\rm h}|=\pi d$, being $d$ the tube diameter, from
Eq.~\ref{G^2tubo and G^graph}, the EPC of a SWNT is:
\begin{equation}
|D^2|=  \frac{a_0^2\sqrt{3}}{2T\pi d} |\tilde D^2|. \label{G^2tubo
and G^graph-bis}
\end{equation}

Using Eq.~\ref{G^2tubo and G^graph-bis} in Eq.~\ref{Dtilde dynamic}
and \ref{Dtilde}, it is possible to obtain the expression for the
non-analytic part of the dynamical matrix within zone folding.
Eq.~\ref{Dtilde dynamic} becomes:
\begin{eqnarray}
\tilde{\Theta}_{q}&=&\frac{A_{\bm \Gamma/\rm \bf
K}a_0^2\sqrt{3}}{2\pi^2d} \sum_{m,n=L,R}|\tilde{D}_{({\rm \bf
K}+{\rm \bf k}'+{\rm \bf q})n,({\rm \bf K}+{\rm
\bf k'})m}|^2  \nonumber \\
&& \int_{-\bar  k}^{\bar k}\frac{f_{{\rm {\bf K+k'}},m}-f_{{\rm
\bf K} + {\rm \bf k'} + {\rm \bf q},n}} {\epsilon_{{\rm \bf
K+k'},m}-\epsilon_{{\rm \bf K}+{\rm \bf k'}+{\rm \bf
q},n}+\hbar\omega_{\rm \bf q}+i\gamma}dk',
 \label{Dtilde dynamic-final}
 \end{eqnarray}
while Eq.~\ref{Dtilde} becomes:
\begin{eqnarray}
\tilde {\Theta}_{q}&=&\frac{A_{\bm \Gamma/\rm \bf
K}a_0^2\sqrt{3}}{2\pi^2d}\sum_{m,n=L,R}|\tilde{D}_{({\rm \bf
K}+{\rm \bf k}'+{\rm \bf q})n,({\rm \bf K}+{\rm \bf
k'})m}|^2\nonumber
\\&&\int_{-\bar k}^{\bar k} \frac{f_{{\rm
{\bf K+k'}},m}-f_{{\rm \bf K} + {\rm \bf k'} + {\rm \bf q},n}}
{\epsilon_{{\rm {\bf K+k'}},m}-\epsilon_{{\rm \bf K}+{\rm \bf
k'}+{\rm \bf q},n}}dk'.
 \label{Dtilde-final}
\end{eqnarray}

Note that, since in zone-folding the electronic states of the tube
are mapped onto those of graphene, all the wavevectors in
Eqs.~\ref{Dtilde dynamic-final},\ref{Dtilde-final} are now
vectorial quantities, and not scalar as in Eqs.~\ref{Dtilde
dynamic},\ref{Dtilde}. The integration is performed along the line
of $ZF_{(n,m)}$ passing through ${\bf K}$. Eqs.~\ref{Dtilde
dynamic-final},\ref{Dtilde-final} also show that the dependence of
$\tilde{\Theta}_{{\rm \bf q}}$ on the translational vector ${\bf
T}$ disappears. \emph{This implies that, for metallic tubes,  any
dependence on tube chirality is lost as well}, and all the
information on tube geometry is given by the $\frac{1}{d}$
dependence only.

As already stated, the relevant EPCs correspond to the graphene
phonons at ${\bm \Gamma}$ and ${\rm \bf K}$, calculated between
electronic states $L$ and $R$ near the Fermi energy (Fig~\ref{2kf}).
In Ref.~\onlinecite{Piscanec2004} it was shown that the only optical
phonons of graphene with a non-negligible EPC are the E$_{2g}$
phonon at ${\bm \Gamma}$ and the A$'_1$ at ${\rm \bf K}$. Thus,
throughout the paper we will consider only the SWNT phonons
corresponding to the  graphene ${\bm \Gamma}$-E$_{2g}$ and to the
${\rm \bf K}$-A$'_1$. For simplicity, we will label them ${\bm
\Gamma}$ ( TO or LO ) and ${\rm \bf K}$. For these phonons, in
Ref.~\onlinecite{Piscanec2004} it is shown that the EPC can be
considered independent from $k$ and assumes the values of
Tab.~\ref{G^2 backscattering e forw.scatt.}
\begin{table}
\begin{center}
\begin{tabular}{c|ccc}
&LO&TO&{\bf K}\\
\hline $|\tilde D_{LR}|^2$=$|\tilde D_{RL}|^2$ & $2\langle
D^2_{\bm \Gamma}\rangle_{\rm F}$&0&
$2\langle D^2_{{\rm \bf K}}\rangle_{\rm F}$\\
$|\tilde D_{LL}|^2$=$|\tilde D_{RR}|^2$ &0&$2\langle D^2_{\bm
\Gamma}\rangle_{\rm F}$&
$2\langle D^2_{{\rm \bf K}}\rangle_{\rm F}$\\
\end{tabular}
\end{center}
\caption{EPC $|\tilde D|^2$ in SWNTs for phonons corresponding to
the graphene ${\bm \Gamma}$-E$_{2g}$ ( LO or TO ) and ${\rm \bf
K}$-A$'_1$ phonons. Electronic states are $L$ and $R$ bands near
the Fermi energy (see Fig.~\ref{2kf}). $\langle D^2_{\bm
\Gamma}\rangle_{\rm F}=45.60$ eV \AA$^{-2}$ and $\langle D^2_{{\rm
\bf K}}\rangle_{\rm F}=92.05$ eV\AA$^{-2}$, from
Refs.~\protect\onlinecite{Piscanec2004,Lazzeri2005}. } \label{G^2
backscattering e forw.scatt.}
\end{table}

The EPC of graphene (and, thus, of SWNTs) are defined in
literature in different ways: (i) the derivative of the hopping
integral with respect to the C-C distance ($\eta$)
\cite{Mahan2003, Park2004, Jiang2004, Perebeinos2005electron},
(ii) the EPC matrix element $D$ of Eq.~\ref{definition of G},
(iii) $D$ times the phonon characteristic length, ($g$)~\cite{
Piscanec2004}. The values of $\eta$, $D$, and $g$ are related by
the following expressions~\cite{Piscanec2004, Lazzeri2005}:
\begin{equation}
|D|^2=9/2\eta^2 {\rm ~~~and~~~}
|g|=|D|\sqrt{\frac{\hbar}{2M\omega_{{\rm \bf q}}}}\label{Eta G e
g}
\end{equation}

The relation between $\eta$ and $D$ is obtained by means of a
first neighbors tight-binding model~\cite{nota02}.

A variety of contrasting EPC values are reported in literature.
Tab~\ref{confronto EPC} compares literature results for the graphene
EPC with those we experimentally measured and calculated in
Refs.~\onlinecite{Piscanec2004},\onlinecite{Lazzeri2006}. This shows
significant discrepancies between previous works and experimental
EPCs, which, on the contrary, are in excellent agreement with our
DFT calculations.
\begin{table}[!tbh]
\begin{center}
\begin{tabular}{l|c|c|c}
\hline
 & $\eta$ [eV\AA$^{-1}]$ &
$|\tilde D_{LR}|^2$ [eV$^2$\AA$^{-2}]$ &
$|\tilde g_{LR}|^2$ [eV$^2$\AA$^{-2}]$ \\
\hline\hline
Mahan\cite{ Mahan2003}              & 4.8-6.2 & 103-172 &  0.092-0.15\\
Park \cite{ Park2004}               &  6   & 164   &  0.146      \\
Jiang\cite{ Jiang2004}              &  1.94   & 17    &  0.015      \\
Perebeinos\cite{ Perebeinos2005electron} &  5.29   & 126    &  0.112      \\
Koswatta\cite{Koswatta2005}           &  6.00   & 162    &  0.144      \\
DFT\cite{ Lazzeri2006}                &  4.57   & 94    &  0.083      \\
Experimental\cite{ Piscanec2004}               &  4.50   & 91    &  0.081 \\
\hline
\end{tabular}
\end{center}
\caption{Comparison between calculated and experimental EPCs for
the ${\bm \Gamma}$ LO phonon, in its equivalent $\eta$, $|D|^2$
and $|g|^2$ expressions. All data in literature were calculated
with tight binding. The relation between $\eta$ and $D$ is
obtained by means of a first neighbors tight-binding model.}
\label{confronto EPC}
\end{table}

\section{Confinement Effects within Static DFPT}

In this Section we present a zone folding technique for the
calculation of pure confinement effects on the phonon frequencies of
SWNTs. All the results and calculations performed in this Section
are obtained within the static approximation

\subsection{Electron and Phonon Zone Folding}

A simple and effective method exists to calculate the SWNTs
electronic structure taking into account confinement only. Indeed,
in Section II-A, it was shown that it is possible to fold the
wavevectors of graphene into those of a $(n,m)$ SWNT. This technique
can be used to obtain the electronic band structure of SWNTs from
that of graphene~\cite{Saito1992,Mintmire1992, SaitoBook,ReichBook}.
We refer to this technique as Electron Zone Folding (EZF). Using EZF
it is possible to obtain the complete electronic structure of any
SWNT doing the calculation just on the 2 atoms unit cell of
graphene. EZF relies on the exact correspondence between the
unrolled SWNT unit cell and graphene, therefore all the effects of
curvature are completely neglected. Despite the simplicity of this
technique, for SWNTs with $d>0.8$ nm, the electronic structure
calculated using EZF does not significantly differ from that
calculated by DFT on the SWNT unit cell~\cite{zolyomi2004}.

The same procedure used to fold electrons can be used to fold
phonons \cite{Jishi1993cpl, Dubay2003, SaitoBook, ReichBook}. We
refer to this as Phonon Zone Folding (PZF). As already mentioned, KA
affect graphene and metallic SWNTs differently, and do not affect
semiconducting SWNTs at all. Thus, PZF is not suitable for the
description of the phonon dispersion of SWNTs close to the
anomalies.

\subsection{Phonon calculation within EZF}

Even if PZF cannot be used for the description of the Kohn
anomalies, this result can still be achieved by means of a folding
approach. Eq.~\ref{DynMat on ph normal coord} shows that the
dynamical matrix of a SWNT depends on its electronic structure
through: (i) the electronic charge density and (ii) an integral of
the electronic states energies over the BZ. Numerically, both
these quantities are obtained by performing discrete sums on a set
of k-points over the tube BZ. These sums can be replaced by sums
on points of $ZF_{(n,m)}$ in the two-dimensional graphene BZ. The
evaluation of $\Theta_q$ within EZF completely neglects the
effects of curvature. Thus, using Eq.~\ref{Dtubo1}, we find that
SWNT dynamical matrices computed within EZF are in the form:
\begin{equation}
\hat{\Theta}_{\rm T}^{\rm flat} = \hat{\Theta}_{\rm G} +
\hat{\Theta}_{\rm conf}
 \label{DtuboPiatto}
\end{equation}

In this Section, we use DFPT to calculate the SWNT dynamical
matrices within EZF. We refer to these calculations as EZF-DFT.

A further increase in the speed of calculations can be obtained
from an appropriate choice of the BZ sampling. The graphene Fermi
surface consists of the two inequivalent points {\rm \bf K} and
{\rm \bf K}'. If the $ZF_{(n,m)}$ lines cross these points, the
tube is metallic, otherwise it is semiconducting. The $\pi$ and
$\pi^*$ bands of graphene near the Fermi points have a conic
shape, so the electronic bands of metallic SWNTs close to {\rm \bf
K} and {\rm \bf K}' are linear, and $\epsilon_{k'}=\pm\beta k'$.
This implies that for metallic tubes the integral in
Eq.~\ref{DynMat on ph normal coord} contains terms that diverge
like $1/k$. These singularities are responsible for the KA onset,
and have to be properly described. The most efficient way to
numerically integrate $1/k$ divergences in one dimension is to use
a logarithmic sampling~\cite{Abramowitz1972}. Thus, to optimize
the efficiency of our calculations, we sample the EZF lines with
evenly-spaced grids of k-points in regions away from the Fermi
surface, and with denser logarithmic grids when crossing {\rm \bf
K} or ${\rm \bf K}'$. Details on the k-point sampling used for
EZF-DFT calculations are reported in Tab.\ref{k-points}.
\begin{table}
\begin{center}
\begin{tabular}{lcrccc}
\hline
tube & Ref.& $T_{\rm e}$ & $n_k$ & M/S& $\ln$  \\
\hline\hline
$(n,n)$ & This work &    4 & 302 & M &y \\
$(n,n)$ & $\shortparallel$ &   77 &  91 & M &y \\
$(n,n)$ & $\shortparallel$&  315 &  71 & M &y \\
$(n,n)$ & $\shortparallel$& 3150 &  51 & M &y \\
$(n,0)$ & $\shortparallel$&    4 & 108 & M &y \\
$(n,0)$ & $\shortparallel$&   77 &  38 & M &y \\
$(n,0)$ & $\shortparallel$&  315 &  28 & M &y \\
$(n,0)$ & $\shortparallel$& 3150 &  18 & M &y \\
$(n,0)$ & $\shortparallel$&    -  &  30 & S &n \\
$(n,m)$ & $\shortparallel$&    -  &  $>\frac{2\pi}{0.12T}$ & S &n \\
$(5,2)$ & $\shortparallel$& 315  &  26 & M &y \\
$(12,3)$ & $\shortparallel$& 315  &  48 & M &y \\
$(16,1)$ & $\shortparallel$& 315  &  23 & M &y \\
$(20,14)$ & $\shortparallel$& 315  &  19 & M &y \\
$(15,6)$  & $\shortparallel$& 315  &  18 & M &y \\
$(14,8)$  & $\shortparallel$& 315  &  47 & M &y \\
$(n,n)$ & Ref.~\onlinecite{ Dubay2003} & $^*$1160 &  57 & M &n \\
$(n,0)$ & Ref.~\onlinecite{ Dubay2003} & $^*$1160 &  33 & M &n \\
$(n,0)$ & Ref.~\onlinecite{ Dubay2003} & $^*$1160 &  13 & S &n \\
$(n,n)$ & Ref.~\onlinecite{ Sanchez-Portal1999} & - &  3 & M &n \\
$(10,0)$ & Ref.~\onlinecite{ Sanchez-Portal1999} & - &  2 & S &n \\
$(8,4)$ & Ref.~\onlinecite{ Sanchez-Portal1999, Reich2001} & - &  1 & S &n \\
$(9,3)$ & Ref.~\onlinecite{ Reich2001} & - &  1 & S &n \\
$(3,3)$ & Ref.~\onlinecite{ Bohnen2004} & $^*1096$ & 129 & M & n\\
$(3,3)$ & Ref.~\onlinecite{ Bohnen2004} & $^*137$ & - & M & n\\
$(n,n)$ & Ref.~\onlinecite{ Connetable2005} & 300 & - & M & y\\
$(5,0)$ & Ref.~\onlinecite{ Connetable2005} & 300 & - & M & y\\

\hline
\end{tabular}
\end{center}
\caption{k-points sampling used in our EZF-DFT calculations and
data from other authors. The number of k-points $n_k$ refers to
the number of non equivalent points the tube BZ. These $n_k$
points fold into $n_kN_c$ points in the graphene BZ, $N_c$ being
the number of unit cells of graphene present in the tube unit
cell. The column $T_{\rm e}$ is the electronic temperature used in
the calculations (in Kelvin). Values marked with $^*$ correspond
to calculations where electrons are smeared using a Hermite-Gauss
instead of the Fermi-Dirac function. The M/S column specifies the
metallic or semiconducting nature of the tubes. The $\ln$ column
specifies if a logarithmic grid is used.} \label{k-points}
\end{table}

\subsubsection{Comparison to other computational techniques}

It is important to discuss how our EZF-DFT results compare to other
computational approaches in literature. These can be divided in the
following categories: PZF, \cite{Jishi1993cpl} tight binding, \cite{
Yu1995el, Yu1995jcp, Menon1996, Barnett2005, Barnett2005ssc}
symmetry adapted methods, \cite{Li2004, Popov2004, Milosevic2005,
Popov2006} and DFT~\cite{ Kurti1998, Sanchez-Portal1999,
Reich2001,Dubay2002, Dubay2003, Bohnen2004, Connetable2005}.

The advantages of EZF-DFT over PZF have already been discussed in
the previous Sections. The advantage of EZF-DFT over conventional
DFT is the possibility of using extremely dense k-point grids and
to converge calculations for a smaller $T_{\rm e}$. This is
evident from a survey of previous DFT calculations. We start
examining the DFT approach used in Refs.~\onlinecite{
Sanchez-Portal1999, Reich2001, Dubay2002, Dubay2003}. These
calculations were performed on the real tube unit cell. Thus the
use of dense k-point grids to converge the phonon frequency at
realistic $T_{\rm e}$ was computationally extremely expensive.
Calculations of Refs.~\onlinecite{Sanchez-Portal1999, Reich2001}
correspond to BZ sampling of at most 3 non equivalent k-points in
the unit cell. The comparison between the number of k-points used
in Refs.~\onlinecite{Sanchez-Portal1999, Reich2001}, and those
used in this and other works, seriously challenges the convergence
of the calculations in Refs.~\onlinecite{Sanchez-Portal1999,
Reich2001}.

Refs.~\onlinecite{Dubay2002, Dubay2003} used much denser k-point
sampling (up to 57) and smeared electrons with an Hermite-Gauss
function at a \emph{fictitious}\cite{Methfessel1989} electronic
temperature $T_f=1160$K. In principle, these calculations should
reproduce the $T_{\rm e}=0$ results, showing PD in metallic tubes.
However, these authors observed no PD, but just a softening of the
phonon frequencies. This happens because their chosen value of
$T_f$ is too large to reach the limit of $T_{\rm e}\to 0$, and
describes the system at a finite, but undefined, value of $T_{\rm
e}$. Calculations at lower $T_{\rm e}$ were prevented by the huge
CPU time requirements.  DFT calculations at lower values of
$T_{\rm e}$ have been reported only for extremely small SWNTs in
Refs.~\onlinecite{ Connetable2005, Bohnen2004}. Here, they
calculated the phonon dispersion of SWNTs with $d\sim4$\AA~ at
$T_{\rm e}$ as low as 137K, showing the presence of KA and PD. The
comparison between the k-points sampling and the electronic
temperature in literature and those used in this paper is reported
in Tab.~\ref{k-points}.

As we will show in the next Sections, the calculations of the
phonon dispersion of large SWNTs at $T_{\rm e}$ as low as 4K can
be be easily done by means of EZF-DFT. This is achieved at the
price of neglecting the effects of curvature, which prevents
EZF-DFT to be used for the study of the radial breathing modes,
and results in inaccuracies in the determination of the phonon
frequencies of those tubes for which curvature effects are not
negligible. This point will be discussed in more depth in Section
VI.

EZF-DFT has also to be compared with tight binding and symmetry
adapted calculations. \emph{Non-ab-initio} techniques, such as
tight binding, offer another way to overcome the extremely large
CPU time required by DFT. Most TB models are unable to describe KA
and PD~\cite{ Yu1995el, Yu1995jcp, Menon1996}. A detailed study of
EPC and $T_{\rm e}$ effects on SWNT phonon within TB was reported
by Barnett et al. \cite{ Barnett2005, Barnett2005ssc}. This study
investigated KA and PD in SWNTs, but the results are in contrast
with those obtained by DFT \cite{Connetable2005}. Another
possibility for the efficient calculations of phonons in SWNTs is
the development of symmetry based force constants methods
\cite{Milosevic2005} and lattice dynamic models based on the screw
symmetry of SWNTs \cite{ Li2004, Popov2004, Popov2006}. So far,
only TB implementations of symmetry adapted techniques were
applied to the calculation of the vibrational properties of SWNTs,
which thus may lack the precision of DFT. The implementation of
screw symmetries into a DFT scheme would be highly desirable, but
would require writing a new dedicated code. On the contrary,
another advantage of EZF-DFT is that it can be performed with the
already existing packages. Finally, it is very important to stress
that all the techniques presented in this Section rely on the
static approximation, and thus neglect dynamic effects.

\subsection{Numerical Results}

In this Section, we use EZF-DFT to perform a systematic study of the
phonon dispersion of SWNTs, paying particular attention to the
description of KA and PD. Calculations are performed within the
static approximation.

\subsubsection{Phonon Dispersions}

Here we use EZF-DFT to calculate phonons of SWNTs of different
diameters and chiralities, focusing on diameter and $T_{\rm e}$
dependence of the KA induced phonon softening.

\begin{figure}[!tp]
\centerline{\includegraphics[width=90mm]{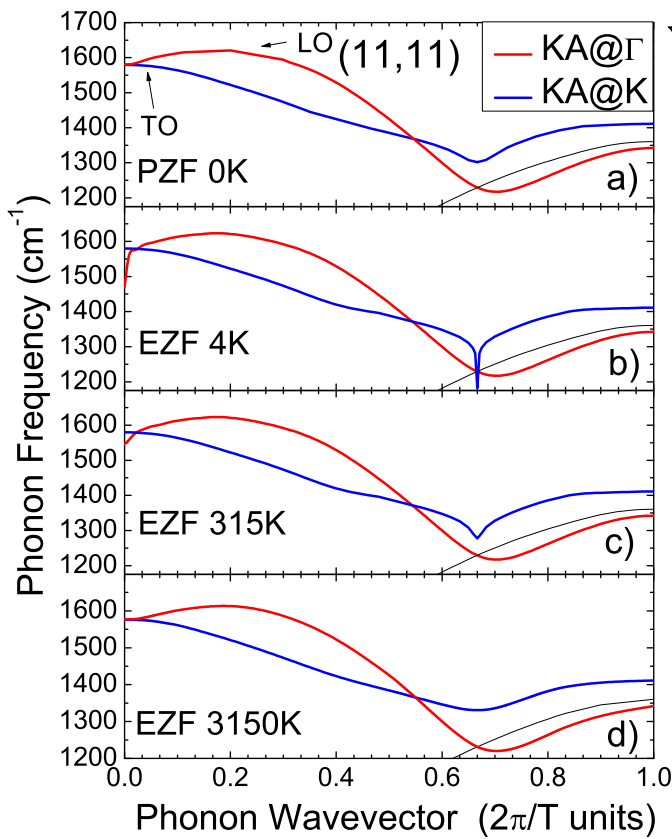}}
\caption{(color on-line) Phonons of the (11,11) metallic SWNT
calculated using (a) Phonon Zone Folding (PZF), (b) Electronic
Zone Folding (EZF-DFT) at $T_{\rm e}=4$K, (c) 315K, and (d) 3150K.
Branches affected by (Red) KA at $\bm \Gamma$, (Blue) KA at {\rm
\bf K}.} \label{Dispersion11-11}
\end{figure}

We start with the comparison of phonon zone folding (PZF) and
EZF-DFT for metallic tubes. Fig.~\ref{Dispersion11-11} compares the
phonons of a metallic (11,11) tube calculated using PZF with those
obtained by EZF-DFT. To improve readability, we only plot phonon
branches corresponding to the lines of $ZF_{(11,11)}$ crossing $\bm
\Gamma$ and {\rm \bf K}. All PZF and EZF-DFT calculations are
performed using DFPT \cite{Baroni2001}. The parameters for PZF
calculations are the same as those used for the calculation of
graphite in Ref.~\onlinecite{ Piscanec2004}.

\begin{figure}[!tp]
\centerline{\includegraphics[width=100mm]{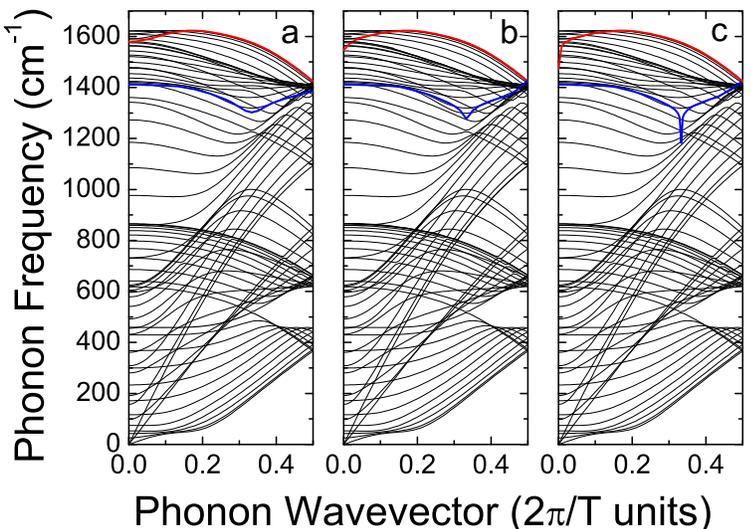}}
\caption{(color on-line) Comparison of the complete phonon
dispersion of a (11,11) tube calculated using (a) PZF, (b) EZF-DFT
at $T_{\rm e}=315$K, (c) EZF-DFT at $T_{\rm e}=4$K. The three
calculations differ only for the branches affected by KA.}
\label{Disp11PZFVsEZF}
\end{figure}

PZF calculations use the phonons of graphene, thus: (i) at ${\bm
\Gamma}$ no LO-TO splitting is observed, and (ii) the kinks and
the phonon softening caused by KA at $\bm \Gamma$ and {\rm \bf K}
are the same as those calculated in Ref.~\onlinecite{
Piscanec2004}. Since the phonon frequencies of graphene converge
at an electronic temperature $T_{\rm e}\sim 3000K$, no temperature
dependence of the phonon frequencies calculated by PZF can be
observed for 0K$<T_{\rm e}<$3000K. EZF-DFT calculations show a
number of fundamental differences with respect to PZF. Fig.
\ref{Dispersion11-11} (b,c,d) show that the phonon softening
induced by KA at $\bm \Gamma$ and {\rm \bf K} strongly depends on
$T_{\rm e}$, with the phonon frequencies decreasing for decreasing
$T_{\rm e}$. Most importantly, we observe that at $\bm \Gamma$
only the LO mode is affected by KA. This causes the splitting of
the LO-TO modes. As already stressed, this splitting cannot be
observed using PZF. Due to the temperature dependence of the LO
mode, the LO-TO splitting increases for decreasing $T_{\rm e}$.
For $T_{\rm e}=3150$K, the KA effects are completely removed, and
the LO-TO splitting is zero. Finally, in Fig.
\ref{Disp11PZFVsEZF}, we compare the whole phonon dispersion of
the (11,11) tube computed using (a) PZF, (b) EZF-DFT at $T_{\rm
e}=4$K and (c) $T_{\rm e}=315$K. Fig. \ref{Disp11PZFVsEZF} shows
that PZF and EZF-DFT differ only for the phonon branches affected
by KA, and are indistinguishable for all other phonon branches.
Thus, in the phonon dispersion of metallic SWNTs only the modes
affected by KA depend on $T_{\rm e}$, and PZF can be safely used
for the calculation of all other phonon modes.

\begin{figure}[!tp]
\centerline{\includegraphics[width=90mm]{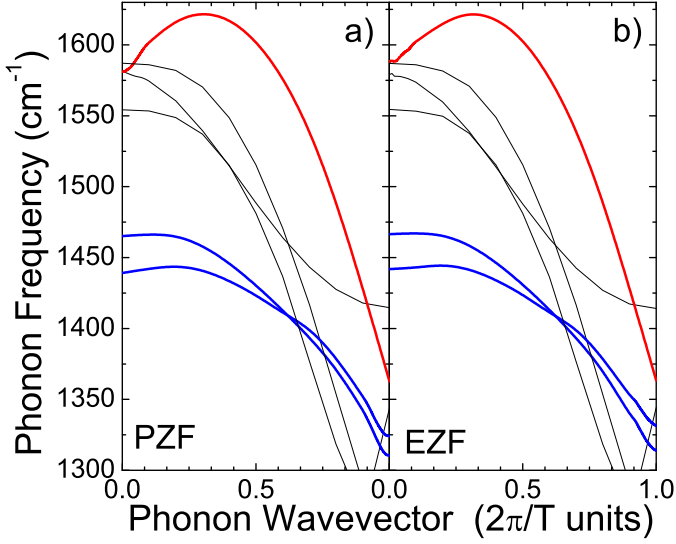}}
\caption{(color on-line) Phonons of the (19,0) tube calculated
using (a) PZF on a q-points grid sampling $ZF_{(19,0)}$, (b)
EZF-DFT. Branches that in graphene and metallic tubes would be
affected by the KA anomaly at $\bm \Gamma$ are in red, while those
closest to the KA at {\rm \bf K} in graphene are in blue.}
\label{Dispersion19-0}
\end{figure}

It is also interesting to compare PZF and EZF-DFT for
semiconducting tubes. Fig.~\ref{Dispersion19-0} compares PZF and
EZF-DFT for a semiconducting (19,0) tube. This has the same
diameter (1.5 nm) as the metallic (11,11) in
Fig.~\ref{Dispersion11-11}. As for Fig.\ref{Dispersion11-11}, in
Fig.\ref{Dispersion19-0} we only plot selected phonon branches, in
particular those corresponding to the line of $ZF_{(19,0)}$
crossing $\bm \Gamma$ and those corresponding to the two lines
that lay most closely to {\rm \bf K}. KA cannot be present in
semiconductors, thus no KA should be observed in the (19,0) tube.
Since $ZF_{(19,0)}$ does not include ${\rm \bf K}$, and the empty
and occupied electronic states are separated by a gap $E_g\sim
0.5$ eV, no temperature effects can be observed for $T_{\rm
e}<E_g/k_{\rm B}\sim 6000$K.

Fig.~\ref{Dispersion19-0} shows that the discrepancies between PZF
and EZF-DFT are extremely reduced with respect to those of the
metallic $(11,11)$ tube, the only noticeable difference being the
absence of LO-TO splitting in the PZF calculations.

\subsection{LO-TO splitting}

Within EZF-DFT, the doubly degenerate $\bm \Gamma-E_{2g}$ phonon
of graphene splits for both metallic and semiconducting tubes.
This splitting is a direct consequence of confinement in SWNTs.

Defining $\omega^{\rm S(M)}_{\rm LO(TO)}$ as the phonon frequency
of the LO (TO) mode at {\rm \bf q}=0 for a semiconducting
(metallic) SWNT, we also observe that $\omega^{\rm M}_{\rm
TO}>\omega^{\rm M}_{\rm LO}$, while $\omega^{\rm S}_{\rm
LO}>\omega^{\rm S}_{\rm TO}$. This is explained as follows. From
Tab.~\ref{G^2 backscattering e forw.scatt.}, the EPCs between $L$
and $R$ bands at $\bm \Gamma$ have a finite value for LO modes,
and are null for TO modes. Since the EPC associated with TO
phonons is null, from Eq.~\ref{Dtilde} we obtain that $\omega^{\rm
graph}_{\rm TO}=\omega^{\rm S}_{\rm TO}=\omega^{\rm M}_{\rm TO}$,
where $\omega^{\rm graph}_{\rm TO}$ is the phonon frequency of the
TO mode at $\bm \Gamma$ for graphene. On the other hand, LO modes
have a finite EPC. Here it is important to remind that what
determines the physical properties related to EPC is not the EPC
\emph{per se}, but the product of the EPC times the electronic
density of states (DOS) (see also Appendix A). Close to the Fermi
energy, for semiconducting SWNTs, the DOS is null, while for
metallic SWNTs it is a constant, and for graphene is zero at Fermi
energy ($\epsilon_{\rm F}$) and increases linearly for
$\epsilon>\epsilon_{\rm F}$ and $\epsilon<\epsilon_{\rm F}$. It
follows that $DOS^{\rm S}_{\epsilon_{\rm F}}<DOS^{\rm
graph}_{\epsilon_{\rm F}}<DOS^{\rm M}_{\epsilon_{\rm F}}$, where
the superscripts $S$, $M$, and $graph$ refer respectively to
semiconducting SWNTs, metallic SWNTs, and graphene, and the
subscript $\epsilon_{\rm F}$ indicates that the DOS is evaluated
for energies $\epsilon$ close to the Fermi energy $\epsilon_{\rm
F}$. From Eq. \ref{Dtilde}, and using the same notation as the TO
modes, we then obtain $\omega_{\rm LO}^{\rm M}<\omega_{\rm
LO}^{\rm graph}<\omega_{\rm LO}^{\rm S}$. Since in graphene we
have $\omega_{\rm LO}^{\rm graph}=\omega_{\rm TO}^{\rm graph}$, it
follows that $\omega_{\rm LO}^{\rm S}>\omega_{\rm TO}^{\rm S}$ and
$\omega_{\rm LO}^{\rm M}<\omega_{\rm TO}^{\rm S}$.

With EZF-DFT, we also find that the atomic displacements of the
modes deriving from the graphene $\bm \Gamma-E_{2g}$ are parallel
or perpendicular to ${\rm \bf q}_{\|}$ within an angle less than
$3.25^{\rm o}$. The effect of confinement is then, for any
chirality, to split the $\bm \Gamma-E_{2g}$ mode of graphene into
almost exactly LO and TO modes.

\subsubsection{Diameter, electronic temperature, chirality and
family dependence of phonon softening}

\begin{figure}[!tp]
\centerline{\includegraphics[width=80mm]{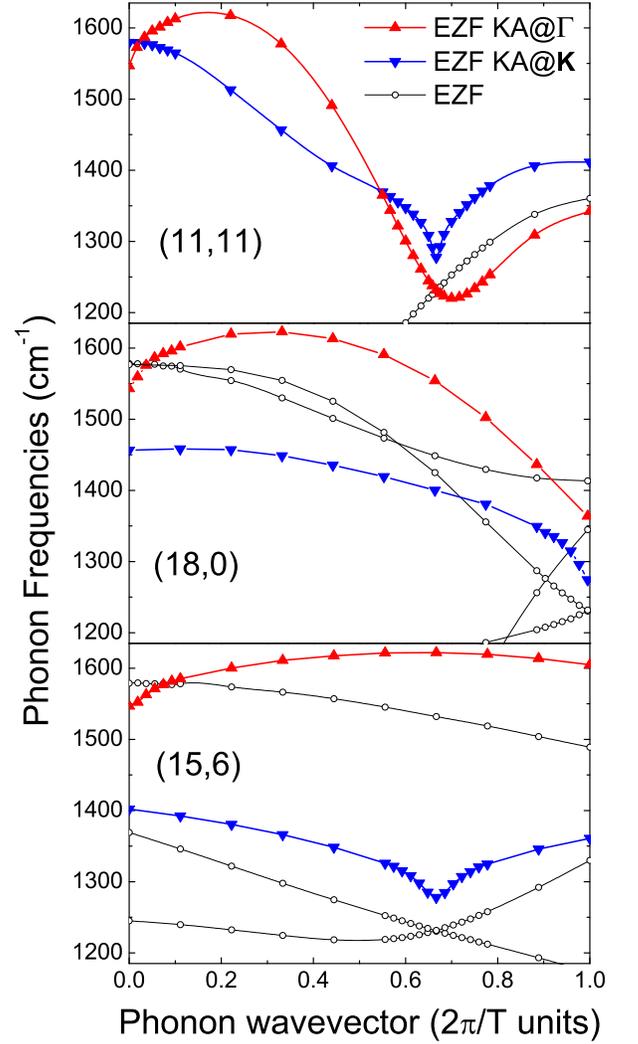}}
\caption{(color on-line) Phonons of the (11,11), (18,0), and
(15,6) tubes calculated using EZF-DFT. All tubes have $d\sim1.5$
nm. Calculations are done at $T_{\rm e}=315$K. Phonon branches
affected by KA at $\bm \Gamma$ are in red, those with KA at {\rm
\bf K} are in blue.} \label{DispersionArmZigChir}
\end{figure}

\begin{figure}[!tp]
\centerline{\includegraphics[width=80mm]{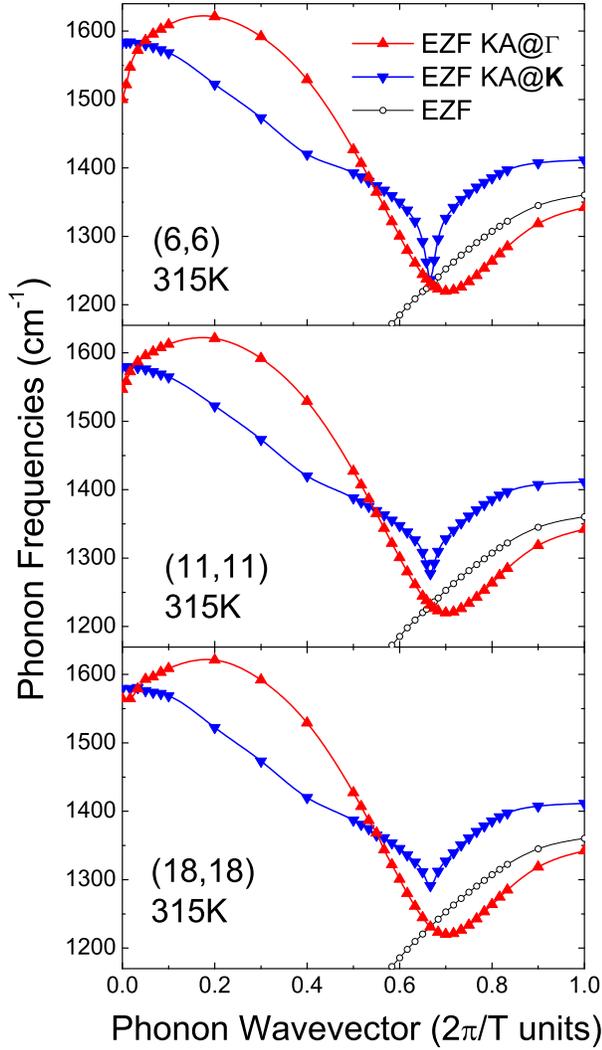}}
\caption{(color on-line) Phonons of the (6,6), (11,11), and
(18,18) SWNTs calculated using EZF-DFT. Tube diameters are 0.8 nm,
1.5 nm, and 2.4 nm respectively. $T_{\rm e}$ is 315K. Phonon
branches affected by KA at $\bm \Gamma$ are in red, while those
with KA at {\rm \bf K} are in blue.} \label{AEZFVsD}
\end{figure}

\begin{figure}[!tp]
\centerline{\includegraphics[width=80mm]{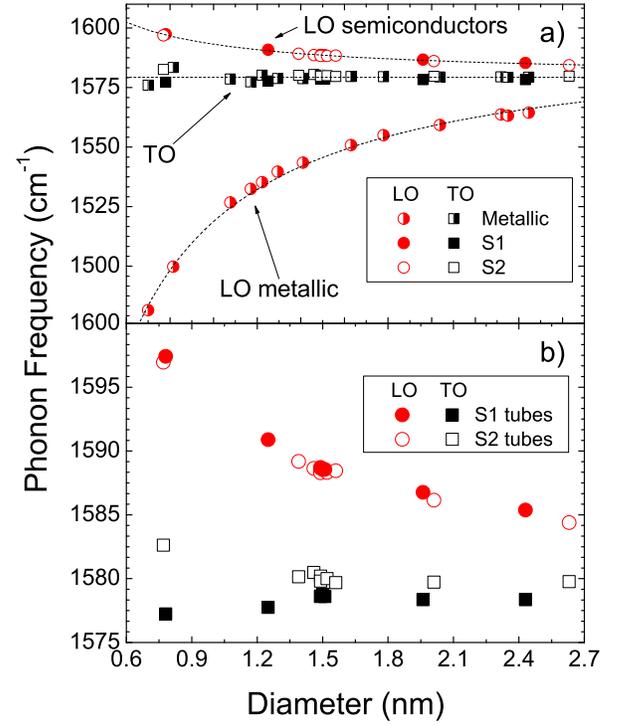}}
\caption{ (color on-line) (a) Frequency of the LO and TO modes
calculated by EZF-DFT at $T_{\rm e}=315$K. The LO phonons are in
red; the TO phonons in black. Different symbols identify metallic
and semiconducting (S1 and S2) tubes. Dashed lines represent the
fit of the EZF-DFT data, and correspond to the plot of
Eq.~\ref{omega T1}. (b) LO and TO phonons of semiconducting tubes
only. For S2 tubes TO phonon is $\sim2$ cm$^{-1}$ higher than in
S1 tubes. No family dependence is observed for the LO phonon}
\label{LO & TO Vs D - tutti i tubi}
\end{figure}

\begin{figure}[!tp] \centerline{\includegraphics[width=80mm]{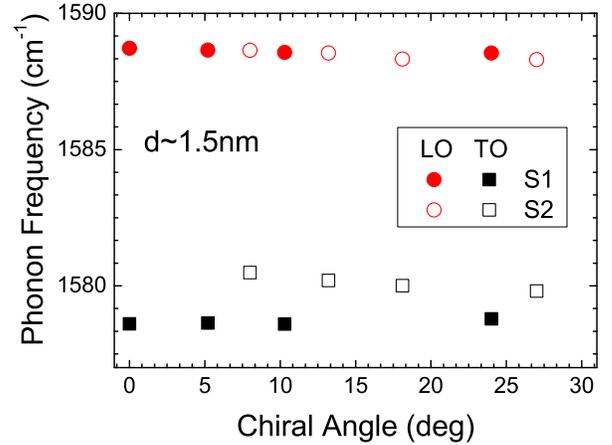}}
\caption{(color on-line) Calculated LO and TO modes obtained by
EZF-DFT for the semiconducting tubes $(19,0), (18,2), (17,3),
(17,4), (16,5), (15,7), (13,9)$, and $(12,10)$, listed by increasing
chiral angle. All tubes have diameter $d\sim 1.5$ nm. Filled symbols
refer to S1 tubes (mod$(n-m,3)=1$); open symbols to S2 tubes
(mod$(n-m,3)=2$). In S2 tubes the TO mode frequency is $\sim
2$cm$^{-1}$ higher than in S1.} \label{SemicD15VsChir}
\end{figure}

We now consider the dependence of the phonon dispersions on SWNT
chirality, diameter and electronic temperature.

Fig.~\ref{DispersionArmZigChir} plots the phonon dispersions at
$T_{\rm e}$=315 K for the (11,11) armchair tube, the (18,0)
zig-zag tube, and the (15,6) chiral  tube. All these tubes are
metallic, and have a similar diameter of 1.5 nm. Note that in
SWNTs, defining $d_{\rm R}=\gcd(2n+m,2m+n)$, ${\rm \bf k}_{\rm
F}=0$ if $(n-m)/3d_{\rm R}$ is not an integer (e.g. the (18,0)
tube), and ${\rm \bf k}_{\rm F}=1/3$, in $2\pi/T$ units, if
$(n-m)/3d_{\rm R}$ is an integer, (e.g. (11,11) and (15,6) tubes)
\cite{ ReichBook}. This explains the different KA positions in
Fig.~\ref{DispersionArmZigChir}. We observe that, even if the
shapes of the phonon dispersions are different, the frequency of
the phonons corresponding to the graphene $\bm \Gamma-E_{2g-LO}$
and {\rm \bf K}$-A'_1$ are the same in all cases, and are
significantly softened with respect to graphene. Thus, the phonon
softening due to KA does not depend on chirality.

On the other hand, Figs.~\ref{AEZFVsD},\ref{Disp11PZFVsEZF}
clearly show that KA-induced phonon-softening strongly depends on
diameter and $T_{\rm e}$. Fig.~\ref{AEZFVsD} compares the phonons
for three different armchair tubes at the same $T_{\rm e}$. It
indicates that the smaller the diameter, the stronger the
softening. Fig.~\ref{Disp11PZFVsEZF} compares phonons for the same
(11,11) tube at three different $T_{\rm e}$. It shows that phonon
softening increases for decreasing $T_{\rm e}$. Note that, as for
Figs.~\ref{Dispersion11-11},\ref{Dispersion19-0},
Figs.~\ref{DispersionArmZigChir},\ref{AEZFVsD} also plot only
branches corresponding to $ZF_{(n,m)}$ crossing $\bm \Gamma$ and
{\rm \bf K}.

Finally, due to the phonon softening induced by the KA, we observe
that in metallic tubes of any chirality the LO branch always crosses
the TO, in contrast with Ref.~\onlinecite{Maultzsch2003prl}, where
an anti-crossing of the branches is predicted for chiral tubes.

Fig.~\ref{LO & TO Vs D - tutti i tubi} plots the LO/TO phonon
frequencies for a variety of chiral and achiral tubes, both
metallic and semiconducting, in the diameter range 0.8-2.7 nm at
$T_{\rm e}$=315K. As expected, we observe no difference in the
phonon frequencies for TO modes of metallic and semiconducting
tubes. Furthermore, confinement does not induce any diameter
dependence of the frequencies of the TO modes.

Due to the KA presence, LO frequencies are very different in
metallic and semiconducting tubes. In metallic tubes there is a
strong diameter dependence, with higher softening for decreasing
diameter. This softening is \emph{entirely} due to confinement.
Chirality does not affect the LO phonons. This is shown in
Fig.~\ref{SemicD15VsChir}, which plots LO and TO modes for all
semiconducting tubes with $d\simeq1.5$ nm as a function of
chirality.

However, even if there is no dependence on the chiral angle for the
LO modes, a family dependence is observed for the TO mode in
semiconducting tubes. Indeed, Fig.~\ref{LO & TO Vs D - tutti i tubi}
(a,b) and Fig. \ref{SemicD15VsChir} show that TO phonons are more
scattered than LO. By labeling S1 SWNTs with mod$(n-m,3)=1$, and S2
SWNTs with mod$(n-m,3)=2$, it is possible to observe that the TO
frequency in S1 tubes is always $\sim2$ cm$^{-1}$ smaller than in S2
tubes. The TO frequency for metallic tubes is in between that of S1
and S2 tubes.
\begin{figure}[!tp]
\centerline{\includegraphics[width=80mm]{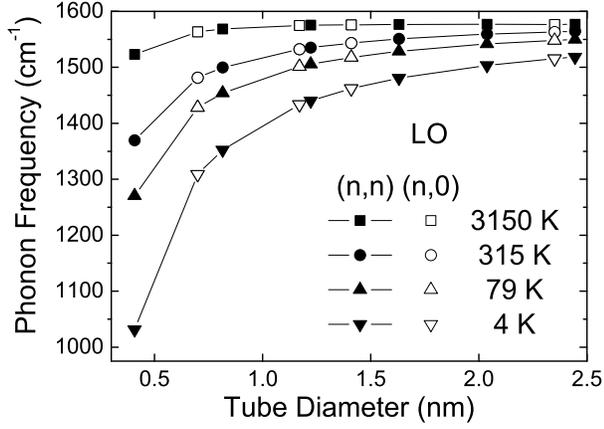}}
\caption{Diameter dependence of LO mode in armchair and metallic
zigzag SWNTs calculated by EZF-DFT for $T_{\rm e}=$4K, 79K,
315K,3150K.} \label{LO Vs T Vs D}
\end{figure}

Fig.~\ref{LO Vs T Vs D} plots the LO phonon of metallic tubes as a
function of $T_{\rm e}$. Since no chirality dependence is observed
for $T_{\rm e}$=315 K, only armchair and zigzag tubes are
considered. The LO softening increases for decreasing $T_{\rm e}$,
and the trend is stronger for smaller diameters.

We stress again, as in Section II-D, that the temperature effects
calculated here with EZF-DFT are related \emph{only} to the
electronic temperature. Therefore, for a direct comparison with
experiments, the data in Fig.~\ref{LO Vs T Vs D} need to be
corrected with the anharmonic effects.

\subsection{Analytic results}

So far we relied on \textit{numerical} DFPT calculations. We now
show that the key results of the previous Section can be derived
and explained by using an~\textit{analytic} model. From
Eqs.~\ref{omega=sqrt(Dq/M)_static},\ref{D=D_an+D_non-an}, we have:
\begin{equation}
\omega_q^2=\frac{\Theta^{\rm
an}_q}{M}+\frac{\tilde{\Theta}_q}{M}\label{w2=w2_an+w2_non-an)}
\end{equation}

KAs are due to the presence of non-analytic terms in the dynamical
matrix and their shape is determined by $\tilde{\Theta}_{{\rm \bf
q}}$, whose expression, within a static approach, is given by
Eq.~\ref{Dtilde-final}. $\tilde{\Theta}_{{\rm \bf q}}$ can be
determined analytically for any \textbf{q} at $T_{\rm e}=0$K, or
at ${\rm \bf q}=\bm \Gamma,{\rm \bf K}$ for any $T_{\rm e}$. This
provides simple formulas for the determination of the
metallic-semiconducting transition associated with the PD and for
the KA description in SWNTs.

\subsubsection{Kohn Anomalies at T=0K and Peierls Distortions}

To determine the KA shape at zero temperature we put $T_{\rm e}=0$
in Eq.~\ref{Dtilde-final}. This gives $f_{{\rm \bf k}}=1$ and
$f_{{\rm \bf k}}=0$ for states respectively below and above the
Fermi energy. Eq.~\ref{Dtilde-final} can then be integrated
analytically to give:
\begin{eqnarray}
\tilde{\Theta}_{{\rm LO}/{\rm \bf K}}&=& \frac{A_{\bm{\Gamma}/{\rm
\bf K}}\sqrt{3} a_0^2 2\langle D^2_{\bm
\Gamma/{\rm \bf K}}\rangle_{\rm F}}{\pi^2 d \beta}\ln|q|+C'_{{\rm LO}/{\rm \bf K}}, \nonumber \\
\tilde{\Theta}_{\rm TO}&=&- \frac{A_{\bm{\Gamma}}\sqrt{3} a_0^2
2\langle D^2_{\bm \Gamma}\rangle_{\rm F}}{\pi^2 d \beta},
\label{DtildeLOTOstatic}
\end{eqnarray}
where $\langle D^2_{\bm \Gamma/{\rm \bf K}}\rangle_{\rm F}$ are
defined in Tab.~\ref{G^2 backscattering e forw.scatt.} and we
assume the $L$ and $R$ bands to be linear with slope $\beta=5.25$
eV \AA. More details on the integration are given in Appendix B.
From Eq.~\ref{DtildeLOTOstatic}, we observe that
$\tilde{\Theta}_{\rm TO}$ does not depend on q. This explains the
absence of KA for the TO mode. On the other hand,
$\tilde{\Theta}_{{\rm LO}/{\rm \bf K}}$ has a logarithmic
dependence on q, which explains the presence of KA in the LO
branch. Using Eq.~\ref{w2=w2_an+w2_non-an)}, we get:
\begin{eqnarray}
\omega^2_{{\rm LO}/{\rm \bf K}}& =& \frac{\alpha_{\bm \Gamma /
{\rm \bf K}}}{d}\ln|q|+C_{{\rm LO}/{\rm \bf K}},
\nonumber \\
\alpha_{\bm \Gamma/{\rm \bf K}}&=&\frac{2\sqrt{3}A_{\bm \Gamma /
{\rm \bf K}} a_0^2 \langle D^2_{\bm {\Gamma / {\rm
K}}}\rangle_{\rm F}} {\pi^2 \beta M} \label{omega_q a T=0},
\end{eqnarray}
where $q$ is measured from $\bm \Gamma$ for the KA at $\bm \Gamma$
and from {\rm \bf K} for the KA at {\rm \bf K}; $\alpha_{{\rm \bf
\Gamma}}=7.89\times 10^{5}$ cm$^{-2}$\AA$^{-1}$; $\alpha_{{\rm \bf
K}}=7.96\times 10^{5}$ cm$^{-2}$\AA$^{-1}$; $C_{{\rm LO}/{\rm \bf
K}}$ account for all the non-divergent terms of the dynamical
matrix.

According to Eq.~\ref{omega_q a T=0}, KA in SWNTs have a
logarithmic shape. For the LO and {\rm \bf K} modes at $T_{\rm
e}=0$K, $\omega^2_{{\rm \bf q}}$ is negative, which gives
imaginary phonon frequencies. This means that the lattice
undergoes a permanent distortion, i.e. a Peierls distortion.

\begin{figure}[!tp]
\centerline{\includegraphics[width=80mm]{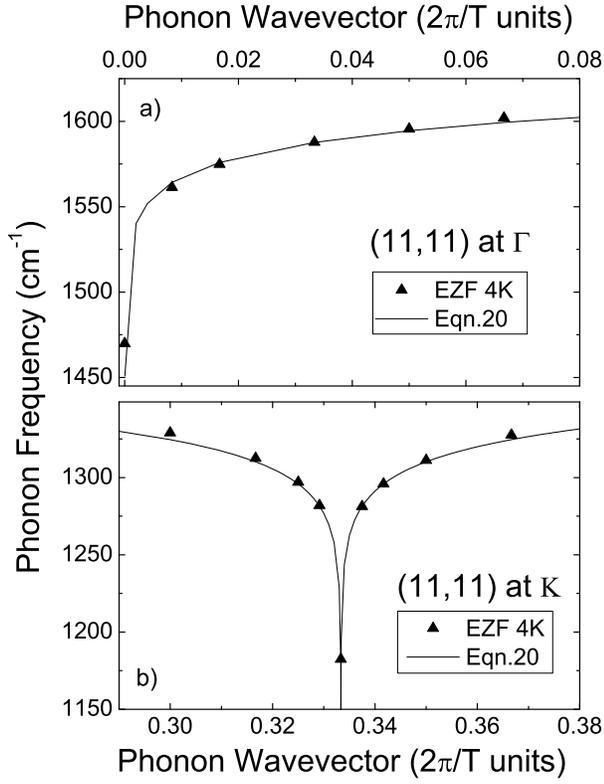}}
\caption{KA at $\bm \Gamma$ and {\rm \bf K} for a (11,11) tube.
Triangles correspond to EZF-DFT at $T_{\rm e}$=4K. The solid line
is the plot of Eq.~\ref{omega_q a T=0}.} \label{KA a T=0}
\end{figure}

Fig.~\ref{KA a T=0} plots Eq.~\ref{omega_q a T=0} for the KA at
$T_{\rm e}=0$K and compares it with EZF-DFT calculations in the
limit $T_{\rm e}\to0$K. Using Eq.~\ref{omega_q a T=0}, and fitting
$C_{{\rm LO}/{\rm \bf K}}$ to the EZF-DFT data, we get:
\begin{eqnarray}
\omega_{\rm LO}^2&=&5.3\times 10^4\ln{q}+2.70\times 10^6 \rm{cm}^{-2},\nonumber\\
\omega_{{\rm \bf K}}^2&=&5.3\times 10^4\ln{q}+1.94\times
10^6\rm{cm}^{-2}\label{KA Gamma e K T=0}
\end{eqnarray}
From Fig.~\ref{KA a T=0}, Eq.~\ref{omega_q a T=0} perfectly
reproduces the phonon dispersions obtained by EZF-DFT.

\subsubsection{Electronic Temperature dependence of Kohn anomalies and Peierls distortion temperature}

We now fix ${\rm \bf q}=\bm \Gamma$ or ${\rm \bf q}={\rm \bf K}$
and study the dependence of the phonon frequencies on $T_{\rm e}$,
which changes the occupation functions $f_{{\rm \bf q},n}$. For
${\rm \bf q}=\bm \Gamma,{\rm \bf K}$, assuming linear $L$ and $R$
bands with slope $\beta=\pm5.52$ eV\AA, remembering that ${\rm \bf
K}+{\rm \bf K}={\rm \bf K}'$, and setting $\epsilon_{\rm F}=0$, we
have:
\begin{eqnarray}
\epsilon_{({\rm \bf K}+{\rm \bf k}'),L}=\epsilon_{({\rm \bf
K}+{\rm \bf k}'+\bm
\Gamma),L}=\epsilon_{({\rm \bf K}+{\rm \bf k}'+{\rm \bf K}),L}={\rm k}'\beta&&\\
\epsilon_{({\rm \bf K}+{\rm \bf k}'),R}=\epsilon_{({\rm \bf
K}+{\rm \bf k}'+ \bm \Gamma),R}=\epsilon_{({\rm \bf K}+{\rm \bf
k}'+{\rm \bf K}),R}=-{\rm k}'\beta&& \label{epsilon pi}
\end{eqnarray}
The occupation factors $f_{{\rm \bf q} ,n}$ depend on ${\rm \bf
q}$ and $n$ only through $\epsilon_n({\rm \bf q})-\epsilon_{\rm
F}$. Defining $x=\beta k'/Tk_{\rm B}$, and
$\theta(x)=1/(1+e^{x})$, Eq.~\ref{Dtilde-final} becomes:
\begin{equation}
\tilde{\Theta}_{\bm \Gamma/ {\rm \bf K}}= \frac{A_{\bm \Gamma/
{\rm \bf K}}2\sqrt{3}a_0^2 \langle D^2_{\bm \Gamma/ {\rm \bf
K}}\rangle_{\rm F}}{\pi^2 d \beta}\int_0^{\frac{\beta \bar
k}{T{\rm k}_B}}\frac{\theta(x)-\theta(-x)}{x}dx,
 \label{Dtilde T non zero}
\end{equation}
where the bands' symmetry allows to replace $\int_{-\frac{\beta
\bar k}{Tk_{\rm B}}}^{\frac{\beta \bar k}{Tk_{\rm B}}}$ with
$2\int_{0}^{\frac{\beta \bar k}{Tk_{\rm B}}}$. This gives:
\begin{equation}
\omega^2_{{\rm LO},{\rm \bf K}}(T)=\frac{\alpha_{\bm \Gamma, {\rm
\bf K}}}{d}\ln\frac{T}{T_0}+ CC_{\bm \Gamma,{\rm \bf
K}}\label{omega T},
\end{equation}
where $\alpha_{{\rm \bf q}}$ is the same as in Eq.\ref{omega_q a
T=0}, $T_0$ is the electronic temperature for which the
contributions of $\tilde{\Theta}_{{\rm \bf q}}$ are null, $CC_{\bm
\Gamma,{\rm \bf K}}$ is the value of $\omega^2_{\bm \Gamma/{\rm
\bf K}}$ at $T_{\rm e}=T_0$, and contains all the contributions
from $\Theta^{\rm an}_{{\rm \bf q}}$. Details on how
Eq.~\ref{omega T} is derived from Eq.~\ref{Dtilde-final} are given
in Appendix C. Eq.~\ref{omega T} can be used to fit
\emph{simultaneously} all the phonon frequencies calculated by
EZF-DFT as a function of $T_{\rm e}$. The fit parameters are
listed in Tab.~\ref{T0 e CC}.
\begin{table}
\begin{center}
\begin{tabular}{c|ccc}
\hline
{\rm \bf q} & $T_0$ (K) & CC (eV$^2$) & CC (cm$^{-2}$) \\
\hline\hline
$\bm \Gamma$ & 9612 & 0.039 & $2.54\times 10^6$ \\
  {\rm \bf K} & 2646 & 0.027 & $1.76\times 10^6$ \\
\hline
\end{tabular}
\end{center}
\caption{values of $T_0$ and $CC_{\bm \Gamma,{\rm \bf K}}$
obtained fitting the EZF-DFT data for tubes calculated at $T_{\rm
e}=4$K, $T_{\rm e}=77$K $T_{\rm e}=315$K $T_{\rm e}=3150$K and for
$d$ in the range 0.8-2.4 nm with Eq.~\ref{omega T}.} \label{T0 e
CC}
\end{table}

The electronic temperature, $T_{\rm PD}$, for which a SWNT
undergoes a Peierls distortion, can be obtained by setting
$\omega_{{\rm \bf q}}$=0 in Eq.~\ref{omega T}, which leads to:
\begin{equation}
T_{\rm PD}=T_0 e^{-\frac{d \cdot CC_{\bm \Gamma,{\rm \bf
K}}}{\alpha_{\bm \Gamma,{\rm \bf K}}}}. \label{PD temperature}
\end{equation}

As an example, Eq.~\ref{PD temperature} applied to the $(6,6)$
tube ($d=0.8$nm) gives $T_{\rm PD}=6.08 \times 10^{-8}$K for the
phonon at $\bm \Gamma$, and $T_{\rm PD}=0.5 \times 10^{-5}$K for
the phonon at {\rm \bf K}. $T_{\rm PD}$ exponentially decreases
with the tube diameter. This implies that, unlike the ultra-small
tubes studied in Refs.~\onlinecite{Connetable2005,Bohnen2004}, for
the SWNTs generally used in experiments, $T_{\rm PD}$ is always
smaller than $10^{-8}$ K.

Eq. \ref{omega T} defines the diameter dependence of the LO phonon
frequency of metallic tubes ($\omega_{\rm LO}^{M}$) at any $T_{\rm
e}$, and can be used to fit the EZF-DFT data of Fig. \ref{LO & TO
Vs D - tutti i tubi}. We observe that all the EZF-DFT data in Fig.
\ref{LO & TO Vs D - tutti i tubi} can be represented by very
simple functions of the tube diameter. Indeed, neglecting the very
small difference observed for the S1 and the S2 tubes, the
frequency of the TO mode of both metallic and semiconducting tubes
($\omega_{\rm TO}$) is diameter independent, while the calculated
frequencies of the LO mode for semiconducting tubes ($\omega_{\rm
LO}^{\rm S}$) are inversely proportional to the tube diameter.
This can be summarized by the following equations:
\begin{equation}
\omega_{\rm LO}^{\rm M}=\eta_{\rm M}+\frac{\nu_{\rm
M}}{d};~~~\omega_{\rm LO}^{\rm S}=\eta_{\rm S}+\frac{\nu_{\rm
S}}{d};~~~\omega_{\rm TO}=\xi, \label{omega T1}
\end{equation}
where $\nu_{\rm M}=-77.33$ cm$^{-1}$ nm, $\eta_{\rm M}=1597$
cm$^{-1}$, $\nu_{\rm S}=13.78$ cm$^{-1}$ nm, and $\eta_{\rm
S}=\xi=1579$ cm$^{-1}$, which perfectly reproduces the phonon
frequency we calculate for the $\bm \Gamma-E_{2g}$ mode of
graphene. The functional form of $\omega_{\rm LO}^{\rm M}$ has
been obtained from Eq.~\ref{omega T} using the relation
$\sqrt{1+x}\sim 1+x/2$. The expression for $\omega_{\rm LO}^{\rm
M}$ is thus completely analytical, but is valid only for $T_{\rm
e}=315K$. On the other hand, the expressions for $\omega_{\rm TO}$
and $\omega_{\rm LO}^{\rm S}$ are empirical, but are valid for any
$T_{\rm e}$. Eqs~\ref{omega T1} are plotted in Fig. \ref{LO & TO
Vs D - tutti i tubi}.

\section{Confinement effects within dynamic DFPT}

In Section II-E, we argued the possible presence of dynamic effects
for one-dimensional metallic systems. To the best of our knowledge,
the dynamic nature of phonons has always been neglected in the SWNT
literature, until we recently pointed this out~\cite{eMRS2006}. In
this Section, we show that dynamic effects are present in metallic
SWNTs, and that they induce significant changes to the KA occurrence
and shape. Therefore, they must be included to explain the
experimental data. In fact, SWNTs are one of the first real
materials for which is a significant difference between static and
dynamic DFPT is detected.

If the static approximation is relaxed, the dynamical matrix is
described by Eq.~\ref{DynMat on ph normal coord dynamic}, and its
non analytic part by Eq.~\ref{Dtilde dynamic}. Assuming $T=0$,
Eq.~\ref{Dtilde dynamic} can be solved by using a procedure similar
to that used to obtain Eq.~\ref{DtildeLOTOstatic}. For the LO
phonon, assuming $\bar{k}\gg|q\pm\frac{\hbar\omega_{{\rm \bf
q}}}{\beta}|$, we get:
\begin{equation}
\tilde{\Theta}_{\rm LO}=\frac{A_{\bm \Gamma}\sqrt{3}a_0^2\langle
D^2_{\bm \Gamma}\rangle_{\rm F}}{\pi^2\beta d}\ln\frac{|\beta
q+\hbar\omega_{\rm LO}||\beta q-\hbar\omega_{\rm
LO}|}{|2\beta\bar{k}|^2}, \label{Dtilde dyn LO}
\end{equation}
while for the TO:
\begin{equation}
\tilde{\Theta}_{\rm TO}=-\frac {A_{\bm \Gamma}\beta
q^2\sqrt{3}a^2_0 2\langle D^2_{\bm \Gamma}\rangle_{\rm F}} {\pi^2
d [\beta^2q^2-(\hbar\omega_{\rm TO})^2]}, \label{Dtilde dyn TO}
\end{equation}
where $\omega_{\rm LO/TO}$ is the phonon frequency of a phonon of
wave vector $q\sim  \Gamma$ belonging to the same branch of the
LO/TO phonon at $q=0$. These equations mark a major departure from
the previous results. $\tilde{\Theta_q}$ in the dynamic approach
is very different from the static case. Note that the static case
(Eq.~\ref{DtildeLOTOstatic}) is immediately recovered setting
$\hbar\omega_{\rm LO/TO}=0$ in Eqs.~\ref{Dtilde dyn
LO},\ref{Dtilde dyn TO}.

The dynamic effects are \emph{qualitatively} very different for LO
and TO phonons. Eq.~\ref{Dtilde dyn LO} diverges for
$q=\hbar\omega_{{\rm \bf q}}/\beta$, and not for $q=0$ as in the
static case. This gives the first major result that the KA for the
LO mode is not at $\Gamma$, contrary to what is predicted by any
static approach, but at $q=\pm \hbar\omega_{\rm LO}/\beta$.

The dynamic effects are even bigger for the TO phonons.
Eq.~\ref{Dtilde dyn TO} shows that $\tilde{\Theta}_{\rm TO}$ is
zero for $q=0$, but diverges to $+\infty$ for $q\to(\hbar
\omega_{\rm TO}/\beta)^-$ and to $-\infty$ for $q\to(\hbar
\omega_{\rm TO}/\beta)^+$. This is quite different from the static
case, where $\tilde{\Theta}_{\rm TO}$ is a negative constant. This
results in a significant TO upshift with respect to the static
case. This can be described by a simple equation, the derivation
of which is given in Appendix D:
\begin{equation}
\omega^{dyn}_{\rm TO}-\omega_{\rm TO}^{stat}\sim\frac{25}{d},
\label{Omega Gamma bs fs bs final TO secVI}
\end{equation}
where the frequency difference is in cm$^{-1}$ and $d$ in nm.

To summarize, the static approach: (i) fails to describe the
position of the KA for the LO mode, (ii) underestimates the TO
phonon frequency, (iii) misses the TO Kohn anomaly at
$q=\hbar\omega_{\rm TO}/\beta$.

The phonon dispersion of metallic SWNTs close to the KAs is
obtained by adding the contributions from the analytic and the
non-analytic part of the dynamical matrix. The contribution of the
non-analytic part is given by the numerical integration of
Eq.~\ref{Dtilde dynamic}. Due to the presence of the phonon energy
in the expression of $\tilde{\Theta}_{q}$, Eq.~\ref{Dtilde
dynamic} has to be solved self-consistently. The contribution from
${\Theta}^{\rm an}_{q}$ is easily obtained from EZF-DFT. The
definitions of $\tilde{\Theta}_{q}$ and $\Theta_q^{\rm an}$ depend
on the value chosen for $\bar k$. In our calculations, we chose
$\bar k$ such that $\bar k \beta=1.0$ eV.
\begin{figure}[!tp]
\centerline{\includegraphics[width=80mm]{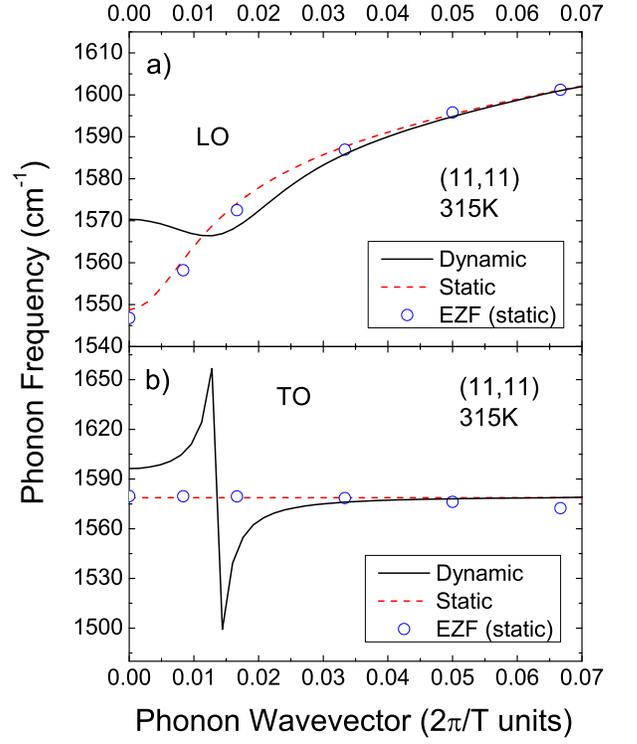}}
\caption{(color on-line) KA for the LO and TO branches of an
(11,11) tube close to $\bm \Gamma$. Phonon frequencies are
calculated by means of static EZF-DFT (open dots), or from the
numerical integration of the dynamical matrix in its static
(dashed line) and dynamic (solid line) expression. $T_{\rm
e}=315$K.} \label{DFT Vs Stat Vs Dyn}
\end{figure}
\begin{figure}[!tp]
\centerline{\includegraphics[width=80mm]{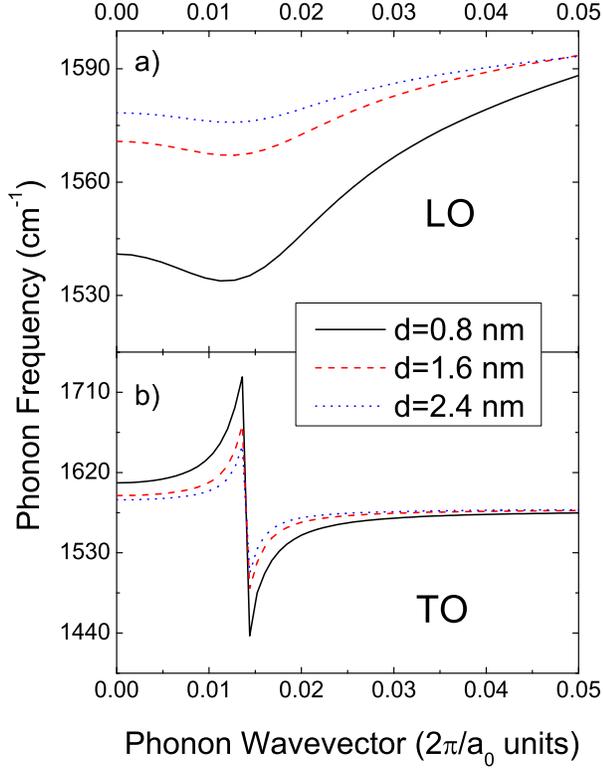}}
\caption{(Color on-line) KA of the LO and TO branches for tubes of
different diameters calculated at $T_{\rm e}=315$K including
dynamical effects.} \label{LO_TO dyn Vs D}
\end{figure}
\begin{figure}[!tp]
\centerline{\includegraphics[width=80mm]{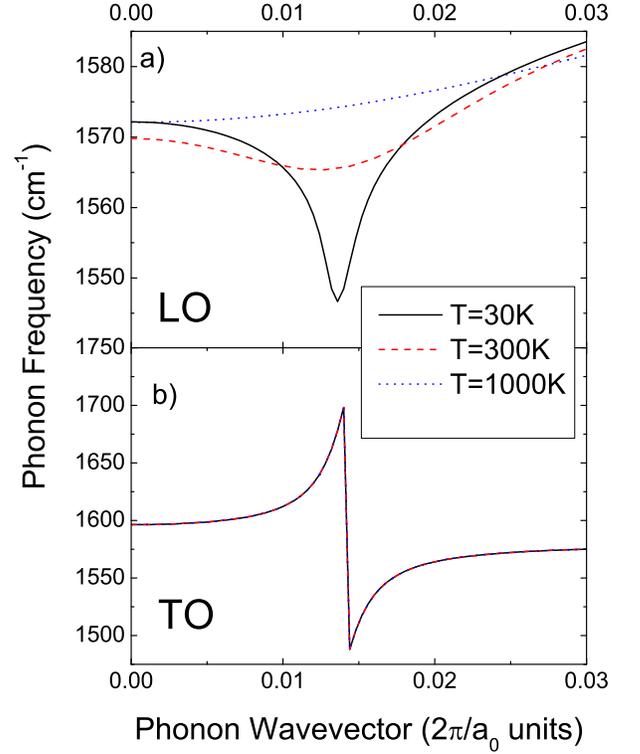}}
\caption{(color on-line) KA of LO and TO branches of tubes with
$d=1.5$ nm calculated for $T_{\rm e}$ ranging from $30\to 1000$K,
including dynamic effects.} \label{LO_TO dyn Vs T}
\end{figure}

Fig.~\ref{DFT Vs Stat Vs Dyn} compares the phonon frequencies of the
LO and TO branches of an (11,11) tube calculated using EZF-DFT with
those obtained from the numerical integration of
Eqs.~\ref{Dtilde-final},\ref{Dtilde dynamic}. Fig.~\ref{DFT Vs Stat
Vs Dyn} shows major differences between the dynamic and the static
approach in the region of the phonon dispersions affected by KA. As
expected, in the dynamic case, the KA in the LO branch is shifted
from $q=0$ to a finite wave-vector. This causes a significant change
close to $\bm \Gamma$. However, Fig.~\ref{DFT Vs Stat Vs Dyn} also
indicates that these changes are confined to a very small BZ region.
Indeed, for ${\rm \bf q}>0.02$ in $2\pi/a_0$ units, the static and
dynamic results are indistinguishable.

It is important to observe that Eq.~\ref{PD temperature}, which
was derived within the static approximation, is also valid in the
dynamic  case. Indeed, for $T_{\rm e}\to 0$, metallic SWNTs
 undergo a PD, meaning that the frequency of
the phonon affected by the KA goes to zero. For $\hbar\omega_{\rm
LO/TO}=0$ the dynamic equations reduces to the static case, so
$T_{\rm PD}$ is the same.

Figs.~\ref{LO_TO dyn Vs D},\ref{LO_TO dyn Vs T} plot respectively
the KA shape in the LO and TO branches as a function of diameter
and $T_{\rm e}$. The general trends observed within the static
approximation are preserved also after the introduction of the
dynamic effects. In particular, the LO phonon softening still
increases for decreasing diameter and $T_{\rm e}$. In addition,
the phonon softening/hardening of the TO branch increases for
decreasing diameters, but has no dependence on $T_{\rm e}$.

\subsection{Dynamic effects in graphene and graphite}

In this Section we have shown that dynamic effects induce huge
modifications in the phonon dispersion of metallic nanotubes close
to the Kohn anomaly. The demonstration of the importance of
dynamic effects in one-dimensional systems opens the discussion on
whether similar effects can be relevant in materials of higher
dimensionality, graphite and graphene in particular. Indeed, it is
possible to prove that dynamic effects can be crucial also for the
correct description of the vibrational properties of
bi-dimensional metallic systems. We have recently demonstrated,
both theoretically and experimentally, that for electrically doped
graphene the Born-Oppenheimer approximation spectacularly
fails~\cite{ Pisana2006, LazzeriMauri2006prl}. In particular the
evolution of the G peak position as a function of gate-bias can be
explained only by taking into account non-adiabatic effects, as
will be reported in details elsewhere~\cite{ Pisana2006,
LazzeriMauri2006prl}.

\section{Curvature Effects}

Curvature effects encompass all the differences between graphene and
SWNTs due to the geometrical distortion of the C-C bonds in the
tubes (Eq.~\ref{Dtubo1}). In common with quantum confinement,
curvature also splits the double degenerate $\bm \Gamma-E_{2g}$
phonon of graphene into two distinct phonons in SWNTs. This split
occurs because the different strengths of the chemical bonds along
the tube axis and the tube circumference result in different phonon
frequencies for modes polarized along these two
directions~\cite{ReichBook, SaitoBook}. The folding of graphene to
give the SWNT cylindrical shape results in: (i) change in atoms
separation (ii) loss of C-C bond planarity, (iii) mixing of $\sigma$
and $\pi$ states, giving a $sp^2$/$sp^3$ character to the chemical
bonds.

When the purely $sp^2$ bonds of graphene are deformed, they assume
a hybrid $sp^2/sp^3$ character. Since $sp^2$ bonds are stiffer
than $sp^3$ bonds, mixing the orbitals results in bond softening.
This mixing is proportional to curvature, so it is minimum for C-C
bonds parallel to the axis, and maximum for bonds oriented along
the circumference~\cite{ReichBook,SaitoBook}.
Since longitudinal modes deform C-C bonds parallel to the axis,
while transverse modes deform C-C bonds along the circumference,
one expects curvature to soften the modes polarized along the tube
circumference more than those polarized along the axis. This
softening should increase with the $sp^2/sp^3$ mixing, and, thus
with the SWNT curvature.

In this Section, we investigate the effects of pure curvature on
the LO-TO splitting. In Eq.~\ref{Dtubo1} the dynamical matrix of a
SWNT ($\hat{\Theta}_{\rm T}$) is written as the sum of the
dynamical matrix of graphene ($\hat{\Theta}_{\rm G}$) and two
additional terms describing the effects of curvature
($\hat{\Theta}_{\rm curv}$) and confinement ($\hat{\Theta}_{\rm
conf}$). EZF-DFT describes the contributions deriving from
graphene and from the confinement effects ($\hat{\Theta}_{\rm
T}^{\rm flat}=\hat{\Theta}_{\rm G}+\hat{\Theta}_{\rm conf}$),
thus:
\begin{equation}
\hat{\Theta}_{\rm curv}=\hat{\Theta}_{\rm T}-\hat{\Theta}_{\rm
G}-\hat{\Theta}_{\rm conf}=\hat{\Theta}_{\rm T}-\hat{\Theta}_{\rm
T}^{\rm flat}. \label{D curv}
\end{equation}
According to Eq.~\ref{D curv}, the effects of pure curvature on a
given SWNT can be obtained as the difference between calculations
performed on the actual SWNT and the results obtained by means of
EZF-DFT.

The frequency dependence on the diameter for the modes derived from
the graphene $\bm \Gamma-E_{2g}$ was calculated by several
authors~\cite{Sanchez-Portal1999, Dubay2002, Dubay2003}. As already
pointed out in Section IV, previous calculations of the LO phonons
in metallic SWNTs are either not converged with respect to the
k-point sampling \cite{Sanchez-Portal1999} or use a fictitious
electronic temperature\cite{Methfessel1989,Dubay2002, Dubay2003}
instead of the real one, and thus do not properly describe the KA
effects. However, for semiconducting SWNTs, KA are not present.
Thus, the DFT results of Refs.~\onlinecite{Dubay2002, Dubay2003} for
LO and TO modes in semiconducting SWNTs are reliable, and can be
used to estimate the curvature effects for LO and TO modes.

To verify the equivalence of the calculations of Refs.
~\onlinecite{Dubay2002,Dubay2003} with the present, the phonon
dispersions of graphene of Ref.~\onlinecite{Dubay2003} are compared
with those of Ref.~\onlinecite{ Piscanec2004} (which is obtained
with the same method and code as the present calculations). Away
from the KA, the only noticeable difference is a 16 cm$^{-1}$ rigid
upshift of the highest optical branches. Thus, the two calculations
are equivalent (after a downshift of 16 cm$^{-1}$ ) and is possible
to extract the effects of pure curvature by comparing the results of
Refs.~\onlinecite{Dubay2002,Dubay2003} with the present EZF-DFT
calculations.

\begin{figure}[!tp]
\centerline{\includegraphics[width=80mm]{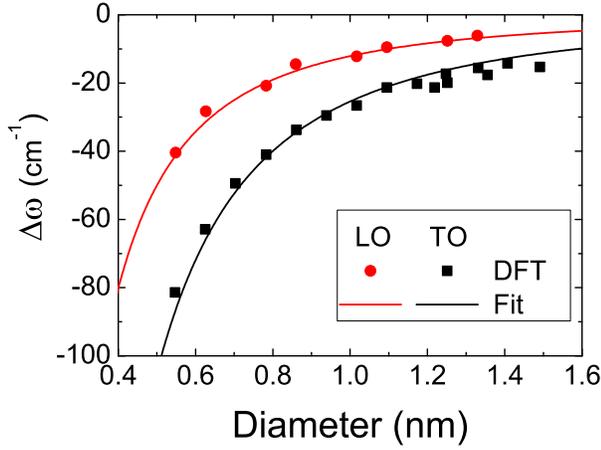}}
\caption{(color on-line) Curvature effects on LO and TO phonons.
Dots represent the difference between the phonon frequencies of
Ref.\onlinecite{Dubay2003} and EZF-DFT of Section IV. (Eq.\ref{omega
T1}). The lines are fit to the data done using Eq.\ref{TOcurv}}
\label{FitKresse}
\end{figure}

Assuming the curvature effects as a perturbation, we define
\begin{equation}
\Delta\omega = \omega_{\rm T} - \omega_{\rm flat}, \label{omega
curv}
\end{equation}
where $\omega_{\rm T}$ is the phonon frequency calculated on an
actual tube and $\omega_{\rm flat}$ is the EZF-DFT result. In
Fig.\ref{FitKresse} we plot $\Delta \omega$ as the difference
between the phonon frequencies $\omega_{\rm T}$ of
Ref.\onlinecite{ Dubay2003} and the $\omega_{\rm flat}$ from
EZF-DFT as represented by Eq. \ref{omega T1}. We consider only
those modes which do not present any dependence on $T_{\rm e}$. In
particular, for the LO mode, we consider only semiconducting
tubes, while for the TO mode, we consider both metallic and
semiconducting tubes. For these modes, it is not necessary to
distinguish between static and dynamic EZF-DFT. From a fit to the
points of Fig.~\ref{FitKresse}:
\begin{eqnarray}
\Delta\omega_{\rm TO}(d)&=&-\frac{\zeta_{\rm TO}}{d^2} \nonumber
\\
\Delta\omega_{\rm LO}(d)&=&-\frac{\zeta_{\rm LO}}{d^2},
\label{TOcurv}
\end{eqnarray}
where $\zeta_{\rm TO}=25.16$ cm$^{-1}$ nm$^2$, $\zeta_{\rm
LO}=12.0$ cm$^{-1}$ nm$^2$, and $d$ is the tube diameter in nm. We
assume that Eq.~\ref{TOcurv} applies also to the LO mode of
metallic tubes.

\begin{figure}[!tp]
\centerline{\includegraphics[width=80mm]{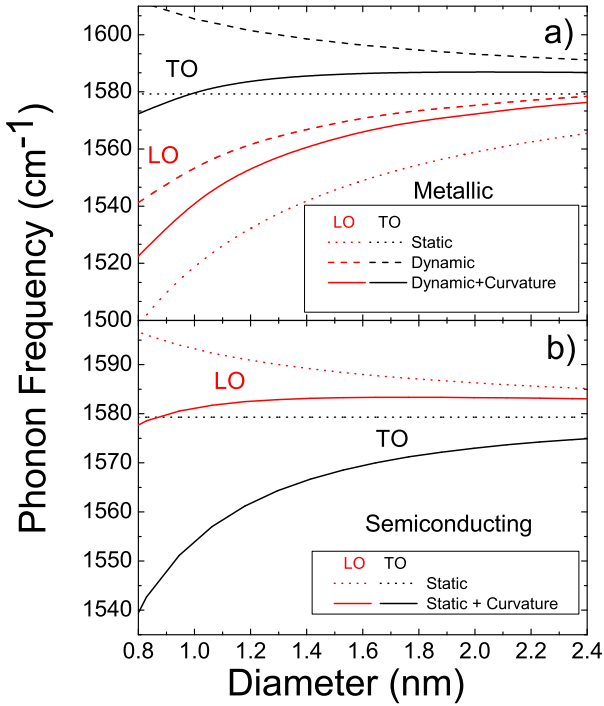}}
\caption{(color on-line) Comparison between LO and TO phonons. (a)
for metallic tubes the phonon frequencies are calculated using
EZF-DFT within the static approximation (dotted line), EZF-DFT
including the dynamic effects (dashed line), and dynamic EZF-DFT
correct with the curvature effects (solid lines). (b) for
semiconducting tubes it is not necessary to distinguish between
static and dynamic calculations; dotted lines represent the
EZF-DFT calculations, solid lines represent EZF-DFT data corrected
for the effects of curvature.} \label{StatcCurvDyn}
\end{figure}

Fig. \ref{StatcCurvDyn} plots the diameter dependence of the LO/TO
phonons as predicted considering only the effects of confinement
(static and dynamic) or after the introduction of the curvature
effects (added using Eq. \ref{TOcurv}). In metallic tubes, the
relaxation of the static approximation introduces a diameter
dependence for the TO phonon, and reduces the softening of the LO
mode. This is a further confirmation that in SWNTs dynamic effects
cannot be neglected. The most notable modification introduced by
curvature is the strong diameter dependence of the TO mode in
semiconducting tubes, which, according to pure confinement, was
predicted to be diameter independent. Interesting, curvature almost
perfectly compensates the effects of confinement for the TO mode in
metallic tubes and for the LO mode in semiconducting tubes,
resulting in almost diameter-independent phonons. Finally, the
effects of curvature in the LO modes of metallic tubes do not
substantially modify the trend due to pure confinement, and result
only in a correction of the phonon frequencies. Fig.
\ref{StatcCurvDyn} is extremely important, and shows that curvature
has the effect of inducing a noticeable phonon softening in small
diameter SWNTs, for both LO and TO modes.

\subsection{LO-TO splitting}

\begin{figure}[!tp]
\centerline{\includegraphics[width=80mm]{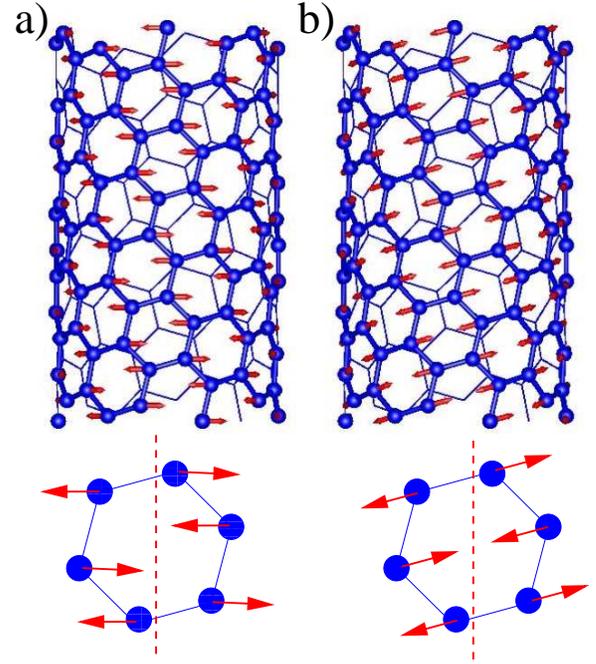}}
\caption{(color on-line) Pattern of atomic displacements
associated to a phonon derived from the graphene $\bm
\Gamma-E_{2g}$ mode (a) as predicted by EZF-DFT and (b) as
calculated in Ref.~\onlinecite{ Reich2001}.} \label{Eigenvectors}
\end{figure}

In Sections IV and V, we observed that quantum confinement has the
effect of splitting the graphene $\bm \Gamma-E_{2g}$ phonon into a
LO and a TO mode. Literature reports indicate that in chiral tubes
curvature effects may induce deviations from this behavior~\cite{
Reich2001, Dubay2003}. In Ref.~\onlinecite{Reich2001}, it was
observed that the eigenvectors of the LO and TO phonons in the
(9,3) tube are aligned to the C-C bonds, and deviate by 16$^{\rm
o}$ from the axial and circumferential directions. However, in the
same paper, it was also reported that for the (8,4) tube the
deviation is reduced to 2$^{\rm o}$. Calculations done in
Ref.~\onlinecite{Dubay2003} show that for the $(12,6)$ tube the
deviation from purely LO and TO eigenvectors is 4$^{\rm o}$. Using
a symmetry adapted TB scheme, calculations performed in
Ref.~\onlinecite{Popov2006} on 300 different SWNTs indicated no
substantial deviation of the atomic displacements from the axial
and circumferential directions. Fig.~\ref{Eigenvectors} shows the
very different predictions given by our EZF-DFT and the
calculations of Ref.~\onlinecite{ Reich2001}.

The orientation of the atomic displacements in phonons derived from
the $\bm \Gamma-E_{2g}$ of graphene can be understood by considering
the combined effects of confinement and curvature. The frequencies
of the $\bm \Gamma-E_{2g}$ modes of graphene are degenerate, and the
corresponding eigenvectors can be oriented along any couple of
in-plane directions. Curvature and confinement both have the effect
of removing the degeneracy. The effect that curvature and
confinement have in orienting the atomic displacements is
proportional to the effect they have in determining the LO-TO
splitting. These contributions can be read from Fig.
\ref{StatcCurvDyn}, by comparing the curves before and after the
introduction of the curvature corrections. In metallic tubes the
effects of confinement are dominant over curvature, thus the modes
derived from graphene $\bm \Gamma-E_{2g}$ are expected to be almost
perfectly LO and TO. In Section IV, and in Tab.~\ref{k-points} we
pointed out that the calculations in Ref.~\onlinecite{Reich2001}
were done using a very limited number of k-points, and thus failed
to describe the effects of confinement. This is also supported by
the results presented in Ref.~\onlinecite{ Dubay2003}. Indeed,
calculations of Ref.~\onlinecite{ Dubay2003} are converged with
respect to the number of k-points, and show that for the (12,6) tube
the modes are almost perfectly LO and TO.

For semiconducting tubes, the effects of curvature are dominant
over the effects of confinement. In this case, the almost perfect LO
and TO character imposed by confinement can be substantially
perturbed by curvature. However, it is important to remember that
the splitting caused by curvature is due to the different force
constants along the tubes axis and circumference. As a
consequence, phonons that present a strong curvature induced
splitting have to be oriented along directions for which the
difference between the force constants are significantly strong.
This means that the atomic displacements have to be
\emph{substantially} aligned along the tube axis and the tube
circumference. This suggests that the deviations from purely
axial and circumferential displacement have to be modest, and that
it should be always possible to classify the modes as
\emph{almost} LO and \emph{almost} TO.

\section{Interpretation of the Raman G band in Nanotubes}

We now show that combining curvature, confinement and dynamic
effects we can interpret the SWNT G peak.

The Raman spectrum of SWNT shows strong features in the 1540-1600
cm$^{-1}$ spectral region\cite{SaitoBook, ReichBook,
Dresselhaus2005, Maultzsch2002prb,
Reich2004pt,Jorio2004pt,acfbook}. This is the same region of the G
peak in in graphene~\cite{Ferrari2006}, graphite
\cite{Tuinstra1970} and amorphous and nanostructured carbons
\cite{Ferrari2000,Ferrari2001}.

\begin{figure}
\centerline{\includegraphics[width=98mm]{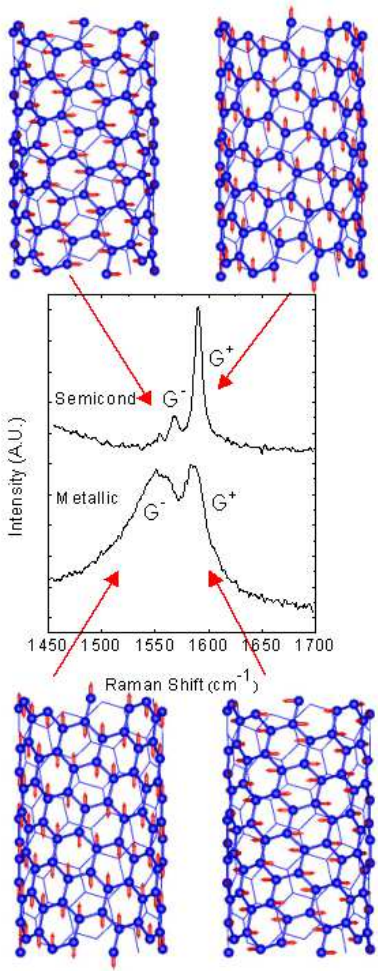}}
\caption{Raman G$^+$ and G$^-$ peaks for semiconducting and
Metallic SWNTs (spectra adapted from Ref.~\onlinecite{
Jorio2004pt}). In semiconducting tubes the two peaks have a
Lorentzian shape and a FWHM of $\sim 12$ cm$^{-1}$. In metallic
tubes the G$^-$ peak is downshifted and much broader. Our
assignment of the spectral features is also indicated.}
\label{G+G-}
\end{figure}

As shown in Fig.~\ref{G+G-}, the G band of SWNTs consists of two
main peaks: G$^{+}$ and G$^{-}$~~\cite{Jorio2002}. In
semiconducting tubes both peaks are sharp, while in metallic tubes
the G$^-$ is broader and
downshifted.~\cite{Pimenta1998,Kataura1999,Rafailov2005,Brown2001,Jorio2002,Jiang2002,Kempa2002,
Maultzsch2003,acfbook,Telg2004,Oron2005,Doorn2005,ReichBook}. For
both metallic and semiconducting tubes, the position of the
G$^{-}$ peak depends on the tube diameter, having lower frequency
for smaller diameters, while the position of the G$^{+}$ peak is
substantially diameter independent~\cite{Jorio2002}. As discussed
in Section VI, curvature affects more strongly the circumferential
modes, thus the G$^{+}$ and G$^{-}$ are commonly assigned to LO
(axial) and TO (circumferential)
modes,respectively.~\cite{Pimenta1998,Kataura1999,Brown2001,Jorio2002,Jiang2002,ReichBook}

Conflicting reports exist on the presence and relative intensity
of the G$^{-}$ band in isolated versus bundled metallic tubes.
Some groups report that the intensity of this peak in isolated
tubes is the same as in bundles,~\cite{Jorio2002,
Oron2005,Doorn2005} while others observe that it is
smaller~\cite{Jiang2002,Maultzsch2003,Telg2004} or even
absent~\cite{Paillet2005}. The downshift and the broadening of the
G$^-$ peak in metallic tubes is commonly attributed to the onset
of a Fano resonance between plasmons and the TO
phonon.~\cite{Kempa2002,Jiang2002,Brown2001,Bose2005,Paillet2005}.
Such phonon-plasmon coupling would either need~\cite{Kempa2002} or
not need~\cite{Brown2001, Bose2005} a finite phonon wavevector for
its activation. The theory of
Refs.~\onlinecite{Bose2005,Brown2001} predicts the phonon-plasmon
peak to be intrinsic in single SWNT, in contrast with
Ref.~\onlinecite{Paillet2005}. On the other hand, the theory in
Ref.~\onlinecite{Kempa2002} requires several tubes ($>$20) in a
bundle in order to observe a significant G$^{-}$ intensity, in
contrast with the experimental observation that bundles with very
few metallic tubes show a significant
G$^{-}$~\cite{Maultzsch2003,Telg2004,Oron2005,Paillet2005,Doorn2005}.
Ref.~\onlinecite{Kempa2002} also predicts a G$^{-}$ upshift with
number of tubes in the bundle, in contrast with
Ref.~\onlinecite{Paillet2005}, which shows a downshift, and with
Refs.~\onlinecite{Jorio2002,Brown2001,Oron2005}, which show that
the G$^{-}$ position depends on the tube diameter and not bundle
size. Finally, the G$^{-}$ position predicted by
Refs.~\onlinecite{Kempa2002,Bose2005} is at least 200 cm$^{-1}$
lower than that measured
~\cite{Pimenta1998,Kataura1999,Rafailov2005,Brown2001,Jorio2002,Jiang2002,
Maultzsch2003,acfbook,ReichBook,Telg2004,Oron2005,Doorn2005}.

Thus, all the proposed theories for phonon-plasmon
coupling~\cite{Kempa2002,Bose2005,Brown2001} are qualitative,
require the guess of several quantities, and fail to predict in a
precise, quantitative, parameter-free way the observed line-shapes
and their diameter dependence.

\begin{figure} \centerline{\includegraphics[width=85mm]{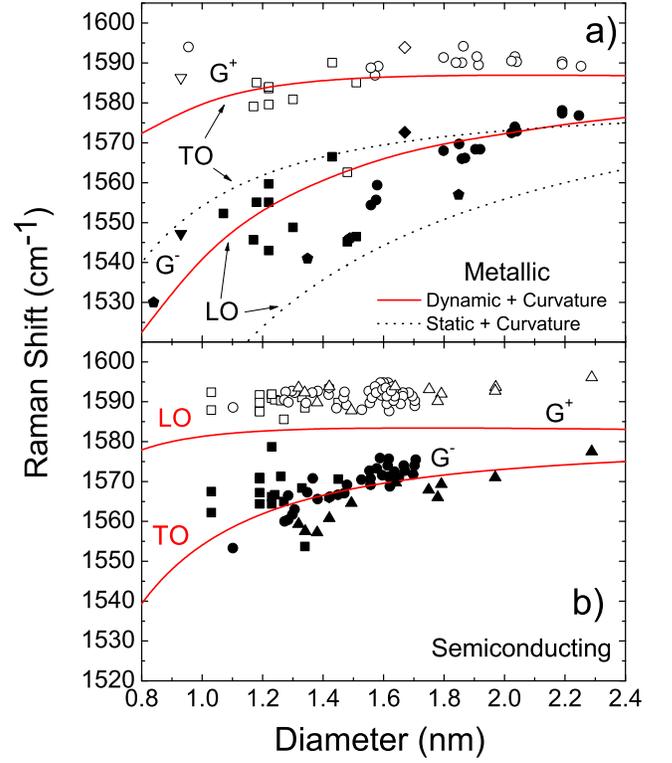}}
\caption{comparison between LO and TO modes of (a) metallic and
(b) semiconducting tubes and the experimental position of the
G$^+$ and G$^-$ peaks of the Raman spectra of SWNTs. The position
of the LO and TO modes is calculated using EZF-DFT and adding the
curvature correction. In panel (a) continuous lines represent the
results of dynamic EZF-DFT calculations, while dotted lines refer
to static EZF. Open and filled symbols represent respectively the
measured position of the G$^+$ and G$^-$ peaks. Data from Oron et
al.\cite{Oron2005} (squares), Jorio et al.\cite{Jorio2002}
(circles), Maultzsch et al.\cite{Maultzsch2003}, (triangles),
Doorn et al.\cite{Doorn2005} (diamonds), Meyier et
al.\cite{Meyer2005} (hexagons), and Brown at al.\cite{Brown2001}
(pentagons) } \label{LO_TO-Vs_G+_G-}
\end{figure}

Phonon symmetry proves that there are only 6 Raman-active mode at
$\Gamma$~\cite{Jorio2003,Jorio2004pt,Dresselhaus2005, Barros2006}:
2 $A$, 2 $E_1$, and 2 $E_2$. The $A$ modes are totally symmetric,
have the highest EPC, and correspond to the LO and TO modes we
examined in the previous Sections. In Fig.\ref{LO_TO-Vs_G+_G-} we
compare the experimentally measured G$^+$ and G$^-$ peaks to the
calculated frequencies of the LO and TO phonons. The plot of the
phonon frequencies corresponds to the results of static and
dynamic EZF-DFT plus the curvature correction $\Delta\omega$.

For semiconducting tubes, our calculations support the G$^+$,
G$^-$ assignment to the LO and TO phonons.

For metallic tubes, our LO and TO positions reproduce the diameter
dependence of the G$^-$ and the G$^+$ peaks respectively. Our
assignment of the G peaks in metallic tubes is thus the opposite
of that in semiconducting. This is emphasized in Fig.\ref{G+G-},
where we represent the correspondence between the Raman peaks and
the LO/TO phonons.

Our calculations prove that the G$^-$ softening can be precisely
predicted by simply considering the coupling of the LO phonon with
single-particle electronic excitations.

Furthermore, in Ref.\onlinecite{Lazzeri2006}, we have shown that
our DFT calculated EPC also account for the measured FWHM.
Vice-versa the experimental FWHM can be used to measure the EPC
\cite{Lazzeri2006}. It is important to remind here that only
single resonance Raman can detect a G$^-$ with large FWHM in
metallic nanotubes\cite{Lazzeri2006}. The same metallic nanotubes
measured in double resonance would have a sharp
G$^-$\cite{Lazzeri2006,Maultzsch2003}.

It is also interesting to consider if Raman scattering can allow
to probe regions very close to the anomalies in Fig.\ref{DFT Vs
Stat Vs Dyn},~\ref{LO_TO dyn Vs D},~\ref{LO_TO dyn Vs T}. This is
not possible in the single resonance approach, since it only
probes $\textbf{q}=0$. This might be possible in double resonant
Raman scattering. In this case the ${\bf q}=0$ rule is relaxed,
and the scattering process involves phonons with $\textbf{q}\neq
0$\cite{ Thomsen2000, Maultzsch2002prb}. However, only phonons
with $q>\hbar\omega_{\bm \Gamma}/\beta$ satisfy the conservation
of energy and momentum in a double resonant process. Fig.~\ref{DFT
Vs Stat Vs Dyn} shows that for such q the anomaly is missed and
the predicted peak position is very similar to the $q=0$ case.
However, crossing from single to double resonant would result in a
significant downshift of the TO mode, as for Eq.\ref{Omega Gamma
bs fs bs final TO secVI}. This is consistent with the experimental
data presented in Ref.~\onlinecite{ Maultzsch2003}.

Further validation of our calculations with experiments is
provided by the analysis of the temperature dependence of the
G$^+$ and G$^-$ positions.

As stressed in the previous Sections, here we do not describe the
frequency shift induced by the presence of anharmonic effects.
However, the effects of anharmonicity are expected to be the same
for the LO and TO phonons. Thus anharmonicity does not affect the
relative position of these two phonons, which can then be altered
only by effects related to $T_{\rm e}$. The temperature dependence
of the $G^+-G^-$ splitting can be investigated using our
analytical model.

\begin{figure}
\centerline{\includegraphics[width=80mm]{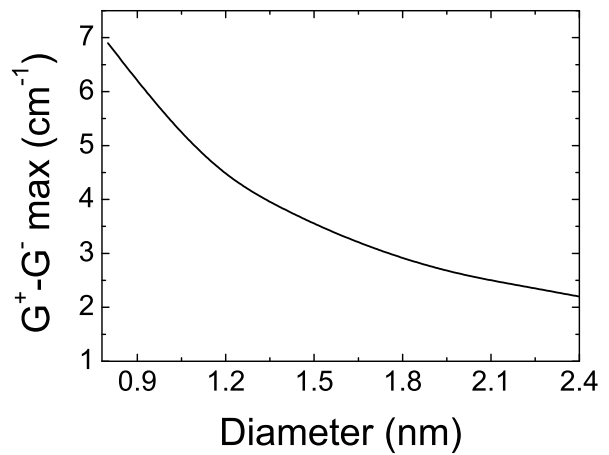}}
\caption{difference between the saturation value at $T_{\rm e}=0$K
and the maximum value at $T_{\rm e} \sim500$K of the frequency
difference of the G$^+$ and the G$^-$ Raman peaks as a function of
the tubes diameter. The values are calculated using dynamic
EZF-DFT and including curvature corrections.} \label{DeltaGMax}
\end{figure}

The difference between the frequency of the LO and TO modes for
metallic and semiconducting tubes of similar diameter $d=1$ nm is
plotted as a function of $T_{\rm e}$ in Fig.\ref{LO dyn Vs T}. For
the semiconducting tube no effects related to $T_{\rm e}$ are
expected, thus the splitting is temperature-independent. For
metallic tubes, the static and the dynamic model give very
different predictions. According to the static model, the
frequency split should have a strong logarithmic dependence on
temperature, Eq.~\ref{omega T}. On the contrary the dynamic model,
obtained by the numerical integration of Eq.~\ref{Dtilde dynamic},
predicts that the splitting (i) has a maximum at $T_{\rm e}\sim
500$K, (ii) it saturates for $T_{\rm e}\to 0$, (iii) for $T_{\rm
e}\to\infty$ it decrease monotonically, and (iv) in the range
$T=0\to1000$K it varies less than 10 cm$^{-1}$. This variation is
function of the tube diameter, and is plotted in
Fig.~\ref{DeltaGMax}. We observe that this value is extremely
small ($\sim 2$ cm$^{-1}$) for large diameters, but rapidly
increases for $d<1.5$ nm.

Experimental data from Ref.\onlinecite{Scardaci2006ump}, obtained
from tubes having $d\sim 1.0$ nm, are in excellent agreement with
the prediction of the dynamical model. This confirms the validity of
our model and the inadequacy of a static approach for the
investigation of the vibrational properties of SWNTs.
\begin{figure}
\centerline{\includegraphics[width=80mm]{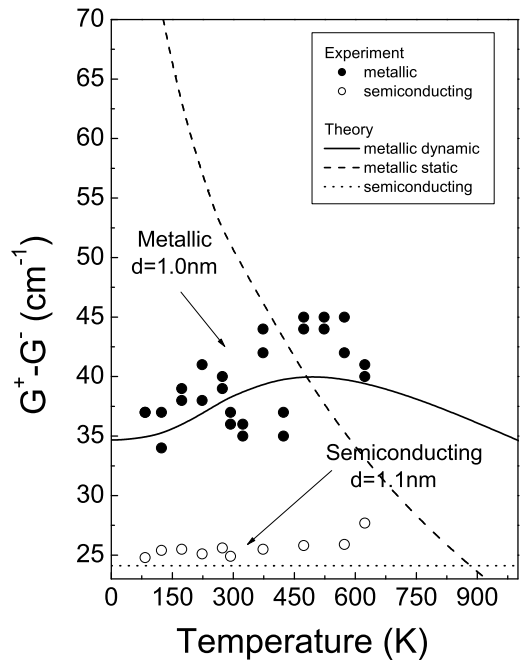}}
\caption{$T_{\rm e}$ dependence of LO-TO splitting for a metallic
and a semiconducting tube, calculated using EZF-DFT and including
curvature corrections. The solid line corresponds to results of
dynamic calculations on a metallic tube; the dashed line
correspond to static calculations. The splitting for a
semiconducting tube is represented by the dotted line.} \label{LO
dyn Vs T}
\end{figure}

\section{Conclusions}

We presented a detailed theoretical investigation of the optical
phonons of SWNTs. We first analyzed the static DFPT approach and
singled out the effects of curvature and confinement. Confinement
plays a major role in shaping SWNT phonons and is often more
relevant than curvature. We presented an electronic zone folding
method allowing the evaluation of confinement effects on
phonon-dispersions and electron-phonon coupling of SWNTs of any
diameter and chirality, and for any electronic temperature. We
investigated Kohn anomalies and Peierls distortions in metallic
SWNTs with diameters up to 3 nm and in a $T_{\rm e}$ range from 4
to 3000 K. We then presented a simple analytic model exactly
accounting for all the static DFPT results. Finally, we proved
that non adiabatic dynamic effects,  beyond the Born-Oppenheimer
approximation, induce significant changes in the occurrence and
shape of Kohn anomalies. We have shown that it is necessary to
consider dynamic effects in order to correctly describe the phonon
dispersions of SWNTs, in contrast to what happens in most other
materials. Finally we discussed the interpretation of the Raman G
peak in SWNTs. Only by combining dynamic, curvature and
confinement effects we can reach an agreement between experiments
and theory. In metallic SWNTs, we assign the G$^+$ and G$^-$ peaks
to the TO (circumferential) and LO (axial) mode, the opposite of
semiconducting SWNTs.

\section{acknowledgements}

Calculations were performed at IDRIS (Orsay, grant 051202) and HPCF
(Cambridge). S.P. acknowledges funding from Pembroke College,
Cambridge, the EU Marie Curie Fellowship IHP-HPMT-CT-2000-00209 and
EU project CANAPE. A. C. F. from EPSRC grant GR/S97613, the Royal
Society and the Leverhulme Trust.

\section{Appendix A}

Here we derive the the relation between the SWNT and graphene EPC
presented in Section III, Eq.~\ref{G^2tubo and G^graph}.

By definition of the EPC (Eq.~\ref{definition of G}):
\begin{equation}
D_{({\rm \bf k}+{\rm \bf q})n,{\rm \bf k}m} = \langle\psi_{{\rm
\bf k}+{\rm \bf q},n}| \frac{\partial\hat H}{\partial \textbf{R}}
\bm\varepsilon_{\textbf{q},\eta} |\psi_{{\rm \bf k},m}\rangle,
\label{definition of G bis}
\end{equation}
where $\hat H$ is the system Hamiltonian, \textbf{R} represents
the atomic coordinates, $|\psi_{\textbf{k},n}\rangle$ is the
electronic state of wavevector \textbf{k} and band $n$, and $\bm
\varepsilon_{\textbf{q},\eta}$ is the polarization of a phonon of
wavevector \textbf{q} and branch $\eta$. We define
$|\psi_{\textbf{k},n}^1\rangle$ and $\bm
\varepsilon_{\textbf{q},\eta}^1$ the wavefunction and polarization
normalized to 1 on the unit cell.  We also define
$|\psi_{\textbf{k},n}^N\rangle$ and $\bm
\varepsilon_{\textbf{q},\eta}^N$ the wavefunctions and the
polarizations normalized to 1 on a supercell composed by $N$ unit
cells:
\begin{eqnarray}
|\psi_{\textbf{k},n}^N\rangle&=&\frac{1}{\sqrt{N}}|\psi_{\textbf{k},n}^1\rangle
\nonumber \\
\bm \varepsilon_{\textbf{q},\eta}^N&=&\frac{1}{\sqrt{N}}\bm
\varepsilon_{\textbf{q},\eta}^1. \label{normalizations}
\end{eqnarray}
The EPC calculated on the supercell is then:
\begin{equation}
D_{(\textbf{k}+\textbf{q})n,\textbf{k},m}^N = \langle
\psi_{\textbf{k}+\textbf{q},n}^N| \frac{\partial\hat H}{\partial
\textbf{R}} \bm\varepsilon_{\textbf{q},\eta}^N |\psi_{
\textbf{k},m}^N\rangle.  \label{definition of G ter}
\end{equation}
This integral is equivalent to the sum of $N$ unit cell integrals:
\begin{eqnarray}
D_{(\textbf{k}+\textbf{q})n,\textbf{k},m}^N &=&
N\langle\frac{\psi_{\textbf{k}+\textbf{q},n}^1}{\sqrt{N}}
|\frac{\partial\hat H}{\partial
\textbf{R}}\frac{\bm\varepsilon_{\textbf{q},\eta}^1}{\sqrt{N}}|
\frac{\psi_{{\rm \bf k},m}^1}{\sqrt{N}}\rangle \nonumber \\
&=&\frac{1}{\sqrt{N}} D_{(\textbf{k}+\textbf{q})n,\textbf{k},m}^1.
\label{definition of G quater}
\end{eqnarray}
Considering a SWNT with a unit-cell containing $N$ graphene unit
cells, Eq.~\ref{G^2tubo and G^graph} is immediately obtained.

Eq.~\ref{definition of G ter} states that the EPC depends on the
supercell choice, as a consequence the EPC \emph{per se} is not a
physical observable. In general, a physical observable is given by
the product of the square of the EPC times an electronic density of
states $\rho$:
\begin{equation}
\rho(\epsilon)=\frac{1}{N_\textbf{k}}\sum_{\textbf{k}n}\delta(\epsilon-\epsilon_n(\textbf{k})),
\end{equation}
where the sum is performed on $N_\textbf{k}$ wavevectors
$\textbf{k}$ and on the bands $n$. $\epsilon_n(\textbf{k})$ is the
energy of $|\psi_{\textbf{k},n}\rangle$. One can easily check that
$\rho$ is proportional to $N$, the size of the super-cell. Thus,
the product $|D|^2\rho$ does not depend on the super-cell chosen
and can correspond to a physical observable.

\section{Appendix B}

We evaluate Eq.~\ref{Dtilde-final} to obtain the shape of the KA
at $T=0$, presented in Section IV-E. For $T=0$K, the occupation
factors $f$ in Eq.~\ref{Dtilde-final} are 0 or 1. For ${\rm \bf
q}\sim\bm\Gamma$ the non-analytic component of the dynamical
matrix is then:
\begin{equation}
\tilde{\Theta}_{{\rm \bf q\sim\bm \Gamma}}=
\frac{A_{\bm{\Gamma}}\sqrt{3} a_0^2}{\pi^2
d}\int_{-\bar{k}}^{\bar{k}} \frac{|\tilde{D}_{({\rm \bf {K+k}'+\bf
q})\pi^*, ({\rm \bf{K+k}'})\pi}|^2} {\epsilon_{({\rm
\bf{K+k}'})\pi} - \epsilon_{({\rm \bf{K+k'+q}})\pi^*}} dk',
\label{Dtilde con T Gamma}
\end{equation}
where $\pi$ and $\pi^*$ indicate bands occupied or empty,
respectively (Fig.~\ref{2kf}). Neglecting the dependence of
$|\tilde D|$ on $k'$ and assuming a linear dispersion for the
electronic bands $\epsilon_{({\rm \bf{K+k'}})\pi}=-\beta k'$ and
$\epsilon_{({\rm \bf{K+k'}})\pi^*}=+\beta k'$, Eq.~\ref{Dtilde con
T Gamma} becomes:
\begin{eqnarray}
\tilde{\Theta}_{{\rm  \bf q\sim \bm \Gamma}}&=&
\frac{A_{\bm{\Gamma}}\sqrt{3} a_0^2}{\pi^2
d}\bigg(|\tilde{D}_{LR}|^2\int_{-\bar{k}}^{-q} \frac{1} {\beta(2k'+q)} dk' \nonumber \\
&+& |\tilde{D}_{LL}|^2\int_{-q}^{0} \frac{1} {-\beta
q} dk'\nonumber \\
&+& |\tilde{D}_{RL}|^2\int_{0}^{+\bar{k}} \frac{1} {-\beta(2k'+q)}
dk'\bigg).\label{Dtilde Gamma bs fs bs}
\end{eqnarray}
Here, the subscripts of $D$ indicate whether we are considering
electronic bands $L$ or $R$ (Fig.~\ref{2kf}). The integrals can be
easily evaluated. Considering
$\ln|2\bar{k}+q|\sim\ln|2\bar{k}|=C_{\bar{k}}$ and $|\tilde
D_{LR}|^2 = |\tilde D_{RL}|^2$, we have:
\begin{equation}
\tilde{\Theta}_{{\rm \bf q\sim \bm \Gamma}}=
\frac{A_{\bm{\Gamma}}\sqrt{3} a_0^2}{\pi^2
d\beta}\bigg[|\tilde{D}_{LR}|^2(\ln|q|-C_{\bar{k}})-|\tilde{D}_{LL}|^2\bigg].
\label{Dtilde Gamma bs fs bs solved bis}
\end{equation}
We now distinguish between the two LO and TO branches at ${\rm \bf
q\sim \bm \Gamma}$. From Tab.~\ref{G^2 backscattering e
forw.scatt.}, we get $|\tilde{D}_{LR}|=2\langle D^2_{\bm
\Gamma}\rangle_{\rm F}$ for the LO branch and 0 for the TO, while
$|\tilde{D}_{LL}|=2\langle D^2_{\bm \Gamma}\rangle_{\rm F}$ for
the TO and 0 for the LO, thus
\begin{eqnarray}
\tilde{\Theta}_{\rm LO}&=& \frac{A_{\bm{\Gamma}}\sqrt{3} a_0^2
2\langle D^2_{\bm
\Gamma}\rangle_{\rm F}}{\pi^2 d \beta}\ln|q|+C'_{\bar{k}} \nonumber \\
\tilde{\Theta}_{\rm TO}&=&- \frac{A_{\bm{\Gamma}}\sqrt{3} a_0^2
2\langle D^2_{\bm \Gamma}\rangle_{\rm F}}{\pi^2 d \beta },
\label{Omega Gamma bs fs bs solved}
\end{eqnarray}
where $\tilde{\Theta}_{\rm LO}$ ( $\tilde{\Theta}_{\rm TO}$ ) is
the non analytic component of the dynamical matrix related to the
LO (TO) branch. $C'_{\bar{k}}=C_{\bar{k}}A_{\bm{\Gamma}}\sqrt{3}
a_0^2 2\langle D^2_{\bm \Gamma}\rangle_{\rm F}/\pi^2 d\beta$
contains all $\bar{k}$ dependent terms. For $q\to 0$
$\tilde{\Theta}_{\rm LO}\to -\infty$, while $\tilde{\Theta}_{\rm
TO}$ is a constant independent on q. This explains the occurrence
of the KA only for the LO branch and its absence for the TO branch
in static DFPT.

The phonon frequencies $\omega$ are obtained by $\omega^2=\Re
e\{(\tilde \Theta + \Theta^{\rm an})/{\rm \emph{M}}\}$, where $M$
is the carbon mass and the analytic component of the dynamical
matrix $\Theta^{\rm an}$ is expected to have a very weak
dependence on \textbf{q}. Concluding, at T=0, for ${\rm \bf
q}\sim\bm\Gamma$, the frequencies of the LO and TO branches are
\begin{eqnarray}
\omega_{\rm LO}^{2}&=&\frac{A_{\bm{\Gamma}}\sqrt{3} a_0^2 2\langle
D^2_{\bm
\Gamma}\rangle_{\rm F}}{\pi^2 d \beta M}\ln|q|+C_{\rm LO}\nonumber \\
&=&\frac{\alpha_{\bm \Gamma}}{d}\ln|q|+C_{\rm LO}
\label{Omega Gamma bs fs bs final LO}, \\
\omega_{\rm TO}^{2}&=&-\frac{A_{\bm{\Gamma}}\sqrt{3} a_0^2
2\langle
D^2_{\bm \Gamma}\rangle_{\rm F}}{\pi^2 d \beta M} + C_{\rm TO} \nonumber \\
&=&-\frac{\alpha_{\bm \Gamma}}{d}+C_{\rm TO}, \label{Omega Gamma
bs fs bs final TO}
\end{eqnarray}
where the constants $C_{\rm LO/TO}$ include all the contributions
from the analytic part of the dynamical matrix $\Theta^{\rm an}$
and all the non-divergent terms of $\tilde{\Theta}$. For the
numerical evaluation of $\alpha_{\bm \Gamma}$ we use: $A_{\bm
\Gamma}=2$, $a_0=2.46$ \AA, $\beta=5.52$ eV \AA, $\langle D^2_{\bm
\Gamma}\rangle_{\rm F}=45.60$ eV$^2$\AA$^-2$, $M=12.01$ a.m.u. The
corresponding equations for the KA at \textbf{K} are obtained in a
completely analogous way.

\section{Appendix C}

We determine the dependence on $T_{\rm e}$ of the KA phonon
frequencies at ${\rm \bf q}={\bm \Gamma}$ and ${\rm \bf q}={\rm
\bf K}$. We define:
\begin{equation}
f(x)=\frac{\theta(x)-\theta(-x)}{x} \label{f(x)}.
\end{equation}
Using Eq.~\ref{omega_q a T=0},
Eq.~\ref{Dtilde T non zero} becomes:
\begin{equation}
\tilde{\Theta}_{\bm \Gamma/ \textbf{K}}= \frac{\alpha_{\bm \Gamma/
\textbf{K}}}{d}\int_0^{\bar{x}} f(x)dx \label{Dtilde T non zero
f(x)},
\end{equation}
where $\bar{x}=\beta \bar{k}/(Tk_{\rm B})$. We observe that
$\lim_{x\to 0}f(x)=0$ and that for $x\to \infty $ $f(x)\sim
-\frac{1}{x}$. Since $\bar{k}$ is chosen in the linear range of
the $\pi$ and $\pi^*$ bands, $\beta k'$ is a fraction of eV.
Moreover, $k_{\rm B}=8.617\times 10^{-5}eV/K$, so the condition
$\beta k' \gg k_{\rm B} T$, corresponding to the $\bar{x} \to
\infty $, is reached for $T$ up to thousands K, well above any
realistic experimental condition. In Eq.~\ref{Dtilde T non zero
f(x)} we replace $f(x)$ with:
\begin{eqnarray}
\tilde{f}(x)&=&f(x) ~~~\rm{for}~~~ x<x_1
\nonumber\\
\tilde{f}(x)&=&-\frac{1}{x} ~~~~\rm{for}~~~x>x_1
\label{ftilde(x)}
\end{eqnarray}
Eq.~\ref{Dtilde T non zero f(x)} becomes:
\begin{equation}
\tilde{\Theta}_{\bm \Gamma/ \textbf{K}}= \frac{\alpha_{\bm \Gamma/
\textbf{K}}}{d}
\bigg[\int_0^{x_1}f(x)dx-\int_{x_1}^{\bar{x}}\frac{1}{x}dx\bigg]
\label{Dtilde T non zero ftilde(x)}
\end{equation}
Defining $F(x)=\int f(x)dx$, we get:
\begin{eqnarray}
\tilde{\Theta}_{\bm \Gamma/ \textbf{K}}&=& \frac{\alpha_{\bm
\Gamma/ \textbf{K}}}{d} \bigg[
F(x_1)-F(0)-\ln{\bar{x}+\ln{x_1}} \bigg]\nonumber \\
&=&\frac{\alpha_{\bm \Gamma/ \textbf{K}}}{d}
\ln\frac{x_0}{\bar{x}} \nonumber \\
&=&\frac{\alpha_{\bm \Gamma/ \textbf{K}}}{d}
\ln\frac{T}{\bar{T_0}}
\label{Dtilde T non zero ftilde(x)bis}
\end{eqnarray}
where $\ln(x_0)=F(x_1)-F(0)+\ln(x_1)$ and
$T_0=\frac{\beta\bar{k}}{k_{\rm B}x_0}$. Considering that the
analytic part of the dynamical matrix
(Eq.~\ref{w2=w2_an+w2_non-an)}) as independent from $T$,
Eq.~\ref{omega T} is then obtained.

\section{Appendix D}

We evaluate the frequency upshift of the TO mode due to dynamic
effects. From Eq.~\ref{Dtilde dyn TO}, we observe that, for q=0,
$\tilde{D}_{\rm TO}=0$. Thus, from Eq.~\ref{w2=w2_an+w2_non-an)},
one obtains that for q=0:
\begin{equation}\label{appD1}
(\omega_{\rm TO}^{\rm dyn})^2=\frac{\Theta^{\rm an}_{\rm TO}}{M}.
\end{equation}
Since the contribution of the phonon energy in the denominator of
$\Theta_{\textbf{q}}$ is negligible for \textbf{k} away from the
Fermi surface, ${\cal D}^{an}_{\textbf{q}}$ is the same in the
static and in the dynamic case. Using
Eqs.~\ref{w2=w2_an+w2_non-an)},~\ref{DtildeLOTOstatic}
and~\ref{omega_q a T=0}, we get:
\begin{equation}\label{appD2}
\frac{{\Theta}^{an}_{\rm TO}}{M}=(\omega_{\rm TO}^{\rm
stat})^2-\frac{\alpha_{\bm \Gamma}}{d}.
\end{equation}
Eqs.~\ref{appD1} and~\ref{appD2} gives:
\begin{equation}\label{appD4}
(\omega_{\rm TO}^{\rm dyn})=\sqrt{(\omega_{\rm TO}^{\rm
stat})^2-\frac{\alpha_{\bm \Gamma}}{d}}.
\end{equation}
Eq.~\ref{Omega Gamma bs fs bs final TO secVI} is finally obtained by
using $\sqrt{1+x}\simeq 1+x/2$ and substituting the numerical value of $\alpha_{\bm \Gamma}$.

\end{document}